\documentclass[aps,pra,showpacs,amssymb,nofootinbib,superscriptaddress,twocolumn,longbibliography]{revtex4-1}

\usepackage{comment}
\usepackage[usenames, dvipsnames]{color}
\usepackage{tikz}
\usetikzlibrary{shapes,backgrounds,fit,decorations.pathreplacing}
\usepackage[T1]{fontenc}
\usepackage[utf8]{inputenc}
\usepackage{bbm, bm, amsbsy, amsthm, amsfonts, amsmath, dsfont, graphicx, epsfig, epstopdf, dsfont, multibib, color,relsize,times}
\usepackage{multirow}
\usepackage{nicematrix}
\usepackage{ulem}

\usepackage[colorlinks,breaklinks]{hyperref}
\makeatletter
\newcommand\org@hypertarget{}
\let\org@hypertarget\hypertarget
\renewcommand\hypertarget[2]{%
  \Hy@raisedlink{\org@hypertarget{#1}{}}#2%
  }
\makeatother
\usepackage[figure,table]{hypcap}
\usepackage{MnSymbol}
\usepackage{enumerate}
\usepackage{float}
\hypersetup{
	bookmarksnumbered,
	pdfstartview={FitH},
	citecolor={darkblue},
	linkcolor={darkblue},
	urlcolor={darkblue},
	pdfpagemode={UseOutlines}}
\definecolor{darkgreen}{RGB}{50,190,50}
\definecolor{darkblue}{RGB}{0,0,190}
\definecolor{darkred}{RGB}{238,0,0}
\definecolor{quantum}{RGB}{83,37,127}
\definecolor{quantumlight}{RGB}{169,146,191}
\usepackage{soul}

\usepackage{paralist} %% Only for lists, can be removed in the end
\usepackage{nicefrac} 
\usepackage{mathtools}
\usepackage{quantikz}

\newcommand{\tr}{\mathrm{Tr}}

\newcommand{\ketbra}[1]{{| #1\rangle \langle #1|}}
\renewcommand{\ket}[1]{|#1\rangle}
\renewcommand{\bra}[1]{\langle #1|}

\newcommand{\Jone}{\mathcal{H}_{\mathcal{J}_1}}
\newcommand{\Jtwo}{\mathcal{H}_{\mathcal{J}_2}}
\newcommand{\J}{\mathcal{H}_{\mathcal{J}}}

\begin{document}

\title{Chiral Symmetries and Multiparticle Entanglement}

\author{Sophia Denker}
\affiliation{Naturwissenschaftlich-Technische Fakult\"at, Universit\"at Siegen, Walter-Flex-Stra{\ss}e 3, 57068 Siegen, Germany}

\author{Satoya Imai}
\affiliation{Institute of Systems and Information Engineering, University of Tsukuba, Tsukuba, Ibaraki 305-8573, Japan}
\affiliation{Center for Artificial Intelligence Research (C-AIR), University of Tsukuba, Tsukuba, Ibaraki 305-8577, Japan}
\affiliation{INO-CNR and LENS, Largo Enrico Fermi, 2, 50125 Firenze, Italy}

\author{Otfried Gühne}
\affiliation{Naturwissenschaftlich-Technische Fakult\"at, Universit\"at Siegen, Walter-Flex-Stra{\ss}e 3, 57068 Siegen, Germany}

\begin{abstract}
Bosons and fermions are defined by their exchange properties and 
the underlying symmetries determine the structure of the corresponding 
state spaces. For two particles there are two possible exchange 
symmetries, resulting in symmetric or antisymmetric behaviour, but 
when exploring multiparticle systems also quantum states with chiral symmetries appear.
In this work we demonstrate that chiral symmetries
lead to extremal forms of quantum entanglement. More precisely, we show 
that subspaces with this symmetry are highly entangled with respect to
the geometric measure of entanglement, leading to observables which
can be useful for entanglement characterization. Along the way, 
we develop a simple method to solve the problem of genuine
multiparticle entanglement for unitarily invariant three-particle states
and use it to identify genuine multipartite entangled states
whose partial transposes with respect to all bipartitions are positive.
Finally, we consider generalizations with less symmetry and discuss 
potential applications. 
\end{abstract}

\date{\today}

\maketitle

%%%%%%%%%%%%%%%%%%%%%%%%%%%%%%%%%%%%%%%%%%%%%%%%%%%%%%%%%%%%%%%%
\textit{Introduction.---}
%%%%%%%%%%%%%%%%%%%%%%%%%%%%%%%%%%%%%%%%%%%%%%%%%%%%%%%%%%%%%%%%
Symmetries are a central concept in physics~\cite{Gross1995,Coleman2010}. 
In the first place, they are central for solving concrete physical problems, where symmetries of the given Lagrange function or Hamilton operator guide the way to find solutions~\cite{Weyl1928, Wigner1931}.
More
generally, symmetries can also occur in the form of permutation symmetries
or gauge symmetries, where they determine the structure of the
state space or the entire theory. For instance, in quantum mechanics the exchange properties of bosons or fermions lead to the notions of symmetric and antisymmetric spaces, and the resulting consequences are still a topic of current research~\cite{WangHazzard2025}.

Entanglement is another central concept in quantum mechanics and is 
known to be useful for quantum teleportation~\cite{BennettBrassardCrepeauJozsaPeresWootters1993}, quantum metrology~\cite{GiovannettiLloydMaccone2004,GiovannettiLloydMaccone2006}, and 
quantum communication and cryptography~\cite{Ekert1991}. Naturally, entanglement has been connected to various types of symmetries, e.g., 
in the construction of hidden variable models~\cite{Werner1989}, the
development of separability criteria~\cite{TothGuehne2009, VollbrechtWerner2001, EggelingWerner2001, Parthasarathy2004} 
 or the characterization of quantum networks~\cite{HansenneXuKraftGuehne2022}.
In the multiparticle scenario, however, there are more than just symmetric and antisymmetric states~\cite{EggelingWerner2001,Ichikawaetal2008,KadiriSivakumar2017,HuberKlepMagronVolcic2022}. 
For instance, there are states that obtain a complex phase after applying 
a cyclic shift operator. This property can be used to characterize so-called chiral spin states~\cite{WenWilczekZee1989}, consequently one may attribute to these states a \textit{chiral} symmetry. Similar to the notion in particle physics, where chiral symmetries describe the transformation
of spinor fields based on their handedness \cite{particle1,particle3,particle2}, the wording \textit{chiral} here connects to the wave function of chiral states, which can be interpreted as a leftwards or rightwards moving current \cite{chiralOp}. 

In quantum information theory, 
chiral symmetries were investigated for example in connection to chiral quantum walks~\cite{ZimborasFaccinKadaWhitfieldLanyonBiamonte2013, LuBiamonteLiLiJohnsonBergholmFaccinZimborasLaflammeBaughLloyd2016} and black holes and chaos \cite{chiralOpnew1, chiralOpnew2, chiralOpnew3}, 
moreover they play an important role in many-body physics 
\cite{Fruchart2016, RoyHarper2017,chiralmb1,chiralmb2,chiralmb3}.

Here, we present cornerstones of entanglement theory in chiral symmetric subspaces. Based on chiral symmetries, we identify highly-entangled subspaces for three particles in arbitrary dimensions, where the entanglement is quantified by the geometric measure~\cite{WeiGoldbart2003,WeinbrennerGuehne2025}. This has two consequences. First, we construct chiral symmetric observables which can be used to verify different classes of entanglement similar to entanglement
witnesses. Second, for the three-qubit case the subspaces can be viewed as parts of a generalized measurement, which can 
be used to estimate Bargmann invariants \cite{Bargmann, Bargmann2,Bargmannrev} and entanglement measures~\cite{LiuKnoerzerWangTura2024} and generalizes maximally entangled measurements (or the so-called swap test) to three parties~\cite{BeckeyPelegríFouldsPearson2022, Bargmanncircuit1,Bargmanncircuit2}. During our analysis we give a 
solution to the problem of genuine multipartite entanglement (GME)
for $U^{\otimes 3}$-invariant states and, as a by-product, we 
uncover $U^{\otimes 3}$-invariant states in local dimension $d\geq 3$, 
which are GME but have a positive partial transpose for all 
bipartitions.
In Appendix~A, we further consider relaxed symmetries and discuss the 
usefulness of these subspaces for applications such as 
quantum data hiding~\cite{HayashiMarkhamMuraoOwariVirmani2006, TerhalDiVincenzoLeung2001}.

%%%%%%%%%%%%%%%%%%%%%%%%%%%%%%%%%%%%%%%%%%%%%%%%%%%%%%%%%%%
\textit{Entanglement and symmetries.---}
%%%%%%%%%%%%%%%%%%%%%%%%%%%%%%%%%%%%%%%%%%%%%%%%%%%%%%%%%%%
Let us give a simple argument why entanglement and 
symmetries are intimately connected. A bipartite state of local dimension $d$ is \textit{symmetric} 
if it is invariant under the exchange of two particles,
$F \ket{\psi} = \ket{\psi}$ and \textit{antisymmetric} 
if it obtains a sign after exchanging two particles, 
$F \ket{\psi} = - \ket{\psi}$. Here, this operation is described by the \textit{flip operator} 
$F = \sum_{i,j=0}^{d-1}\ket{ij}\bra{ji}$, so that $F \ket{ab} = \ket{ba}$. 
Using this, the projectors onto the 
two-particle symmetric and antisymmetric subspace are
\begin{align}
    \Pi_\mathcal{S} = \frac{\mathds{1}+F}{2},
    \quad
    \Pi_\mathcal{A} =\frac{\mathds{1}-F}{2}.
\end{align}
Clearly, $\Pi_\mathcal{S} \Pi_\mathcal{A} = \Pi_\mathcal{A} \Pi_\mathcal{S} = 0$ and $\Pi_\mathcal{S} + \Pi_\mathcal{A}  = \mathds{1}$. In other 
words, the total space is a direct sum of the symmetric
and antisymmetric space, $\mathcal{H} = 
\mathcal{H}_{\mathcal{S}}\oplus\mathcal{H}_{\mathcal{A}}$.
Considering the overlap of a product vector with the symmetric subspace,
\begin{align}
\label{eq-symmtrick}
    \bra{ab}\Pi_\mathcal{S}\ket{ab} &= \frac{1}{2}(1+\bra{ab}F\ket{ab})
    = \frac{1}{2}(1+|\braket{a}{b}|^2) \geq \frac{1}{2},
\end{align}
we find that this overlap is bounded from below by $\nicefrac{1}{2}$. 
Consequently, the overlap of a product state with the 
antisymmetric space is bounded from above by $\nicefrac{1}{2}$.
Then, for all antisymmetric states $\ket{\psi}\in \mathcal{H}_\mathcal{A}$, 
it holds that
\begin{align}
G(\ket{\psi}) = 1 - \max_{\ket{ab}} |\braket{ab}{\psi}|^2 \geq \frac{1}{2},
\end{align}
where $\Lambda^2(\ket{\psi})=\max_{\ket{ab}} |\braket{ab}{\psi}|^2$ denotes 
the largest overlap of $\ket{\psi}$ with a product state $\ket{ab}$. 
The quantity $G(\ket{\psi})$ is known as the \textit{geometric measure 
of entanglement} and can be directly extended to the multiparticle case~\cite{WeiGoldbart2003, WeinbrennerGuehne2025}.
This result implies that the antisymmetric
subspace is an entangled subspace~\cite{Parthasarathy2004}, but the quantitative statement is, of course, stronger. 

\textit{Three qubits.---}
In the three-qubit case the Hilbert space $\mathcal{H} = 
\mathcal{H}_{\mathcal{S}}\oplus\mathcal{H}_{\mathcal{J}}\oplus\mathcal{H}_{\Bar{\mathcal{J}}}$
splits into three subspaces with different symmetries~\cite{Harrow2005, BaconChuangHarrow2006}.
Here, there is no antisymmetric subspace and the symmetric subspace $\mathcal{H}_\mathcal{S}$ contains all states 
that are invariant under the exchange of two particles.
The projector on the symmetric subspace reads 
\begin{align}
\label{eq-pisindequal2}
\Pi_\mathcal{S} = \frac{\mathds{1}+T+T^2}{3},
\end{align}
where $T = \sum_{i,j,k=0}^{d}\ket{kij}\bra{ijk}$, here with $d=2$ is the \textit{cyclic translation operator} (obeying
$
T\ket{abc} = \ket{cab}
$),
and applying $T^2=T^\dagger$ performs an anticyclic translation. 
In fact, $T$ has four eigenvalues $+1$ corresponding to the four basis 
vectors of the symmetric subspace, and two further eigenvalues, 
$\omega = e^{\frac{2\pi \mathbf{i}}{3}}$ and $\omega^2 = e^{\frac{-2\pi \mathbf{i}}{3}} 
= \omega^*$, each twofold degenerate, where the eigenvectors corresponding 
to the eigenvalue $\omega$ are 
\begin{align}
    \ket{\phi_1} &= \frac{1}{\sqrt{3}}(\ket{001} + \omega \ket{010} + \omega^2 \ket{100}), 
    \nonumber \\
    \ket{\phi_2} &= \frac{1}{\sqrt{3}}(\ket{110} + \omega \ket{101} + \omega^2 \ket{011})
\end{align}
and define a basis of the \textit{chiral symmetric} subspace 
$\mathcal{H}_{\mathcal{J}}$~\cite{WenWilczekZee1989, DAlessandroHartwig2021}.
The eigenvectors to the eigenvalue $\omega^2$ are given 
by the complex conjugate of $\ket{\phi_1}$ and $\ket{\phi_2}$ and span 
the \textit{antichiral symmetric} subspace $\mathcal{H}_{\Bar{\mathcal{J}}}$. Hence, we have $T\ket{\phi_i}=\omega\ket{\phi_i}$ and $T\ket{\phi_i^*}=\omega^2\ket{\phi_i^*}$ for $i=1,2$, characterizing the chiral symmetry.
The projectors onto the chiral and the antichiral subspace, respectively, are
\begin{align}
\Pi_\mathcal{J} = \frac{\mathds{1}+\omega^2 T + \omega T^2}{3},
\quad
\Pi_{\Bar{\mathcal{J}}} = \frac{\mathds{1}+\omega T + \omega^2 T^2}{3}.
\label{eq:chiralprojector}
\end{align}
By defining an analoguous cyclic permutation operator $T$ on $N$ parties, we can also identify chiral subspaces in $N$-partite systems as we discuss in Appendix~B.
In Ref.~\cite{SteinbergGuehne2024}, based on a
numerical iteration and without any reference to symmetries, a two-dimensional space was found, where all vectors appear to
be maximally entangled in terms of the geometric measure. In fact, this space is, up to some relabeling, the chiral subspace $\mathcal{H}_{\mathcal{J}}$.
With the proper formulation in terms of symmetries we can characterize the entanglement rigorously:

\textbf{Observation 1.}\label{obs1} \textit{For three qubits, any pure state in the chiral space 
$\mathcal{H}_{\mathcal{J}}$  (or, equivalently, the antichiral 
space $\mathcal{H_{\Bar{\mathcal{J}}}}$) has a geometric measure 
$G(\ket{\psi}) = \nicefrac{5}{9}.$ All states in the direct sum 
$\mathcal{H}_{\mathcal{J}}\oplus \mathcal{H}_{\Bar{\mathcal{J}}}$
have an entanglement of at least $G(\ket{\psi})\geq\nicefrac{1}{4}.$}

The proof of this theorem follows the strategy presented for the 
bipartite case, see Eq.~(\ref{eq-symmtrick}): One considers the 
orthogonal spaces, and shows that the overlap of any product vector 
cannot be small. In particular throughout the proof one can make use of recent results connected to Bargmann invariants \cite{FernandesWagneretal,ChaBarkmann,ChaBarkmann2}. Details are given in Appendix~C.

This result is remarkable in several aspects. First, it is known
that among pure states of three qubits the $W$ state, defined as $\ket{W}\propto \ket{001}+\ket{010}+\ket{100}$ has the maximal geometric
entanglement,  $G(\ket{W})=\nicefrac{5}{9}$~\cite{TamaryanWeiPark2009}, 
but  the observation demonstrates that there is an entire two-dimensional
space where all states are maximally entangled. Note that this is no contradiction to the fact that the $W$ state maximizes the geometric measure, since
all states in this subspace are local unitary equivalent to the $W$ state~\cite{DerksenFriedlandLimWang2017, AlbertiniDAlessandro2021, LiuKnoerzerWangTura2024,Stojanovic2026}.

Second, the projectors $\Pi_\mathcal{S}$, $\Pi_\mathcal{J}$ and 
$\Pi_{\Bar{\mathcal{J}}}$ may also be seen as parts of a generalized
measurement with highly entangled effects. In fact, they can be used to 
measure invariants like $\tr(\varrho^3)$, connected to Bargmann invariants \cite{Bargmann,Bargmann2,Bargmannrev} or entanglement parameters 
like the recently introduced generalized concentratable 
entanglement on several copies
of a state~\cite{LiuKnoerzerWangTura2024}. In this sense, this measurement may be seen as a generalization
of Bell state measurements to three qubits~\cite{BeckeyPelegríFouldsPearson2022, Bargmanncircuit1,Bargmanncircuit2}. We discuss more details in Appendix~D.

Third, our formulation is directly connected
to the identification of the maximally entangled 
four-qubit state~\cite{DerksenFriedlandLimWang2017}.
Given, that any superposition
$\ket{\psi} = \alpha \ket{\phi_1} + \beta \ket{\phi_2}$ has the same 
maximal overlap 
$\Lambda^2(\ket{\psi}) = \max_{\ket{abc}} |\braket{abc}{\psi}|^2 = \nicefrac{4}{9}$,
consider
the four-qubit superposition
$\ket{M} = \nicefrac{1}{\sqrt{2}} (\ket{\phi_1}\otimes\ket{1}+\ket{\phi_2}\otimes\ket{0})$~\cite{HiguchiSudbery2000, BrierleyHiguchi2007}. For this state, we find $\Lambda^2(\ket{M}) = \nicefrac{2}{9}$, since $\max_{\ket{abcd}} |\braket{abcd}{M}|^2 = \nicefrac{1}{2} \times \Lambda^2(\alpha \ket{\phi_1} + \beta \ket{\phi_2})$, where $\alpha = \braket{d}{1}$ and $\beta = \braket{d}{0}$, leading to $G(\ket{M})=\nicefrac{7}{9}$. 

In order to study further consequences and generalizations, we consider the following observable, 
\begin{align}
W_\varepsilon 
= 
\sum_{i,j,k=1}^3\varepsilon_{ijk}\sigma_i\otimes\sigma_j\otimes\sigma_k
= 
2\sqrt{3} (\Pi_{\Bar{\mathcal{J}}}-\Pi_\mathcal{J}),
\label{eq:XA2}
\end{align}
where $\varepsilon_{ijk}$ is the Levi-Civita symbol
and $\sigma_i$ are the traceless Pauli matrices~\cite{ImaiTothGuehne2024}.
The relation to the spaces $\Pi_\mathcal{J}$ and $\Pi_{\Bar{\mathcal{J}}}$
follows directly from the fact that $W_\varepsilon$ coincides with the chirality operator \cite{chiralOp}. 

This observable can witness different classes of entanglement. Recall that a three-particle state is fully separable if
it is a convex combination of fully separable product states of the form
$\ket{\varphi^\mathrm{fs}} = \ket{abc}.$ Moreover, a pure state is biseparable (with
respect to the partition $A|BC$) if it is of the form $\ket{\varphi^\mathrm{bs}} = \ket{\eta_A}\ket{\mu_{BC}}$ and a mixed state is biseparable, if it
is a mixture of pure biseparable states (potentially, for different 
bipartitions)~\cite{AcínBrussLewensteinSanpera2001, GuehneToth2009}.
Finally, states which are not biseparable are GME.

For fully separable states the expectation value of $W_\varepsilon$
is bounded by $|\langle W_\varepsilon\rangle| = 
|\tr(\varrho W_\varepsilon)|\leq 1$  and for all biseparable states
it holds $|\langle W_\varepsilon\rangle|\leq 2$~\cite{ImaiTothGuehne2024, chiralOp}.
This makes the observable $W_\varepsilon$ useful for entanglement detection. In a
similar fashion, the observable 
$\mathcal{W}= \nicefrac{4}{9} \openone -\Pi_\mathcal{J}$ is a witness
excluding full separability, improving standard entanglement witnesses
of the type $\mathcal{W}= \nicefrac{4}{9} \openone -\ketbra{W}$~\cite{WeiGoldbart2003,Rico}.

It further turns out that the observable $W_\varepsilon$ is invariant 
under unitary transformations of the form $U^{\otimes 3} W_\varepsilon (U^{\otimes 3})^\dagger=W_\varepsilon$
\cite{ImaiTothGuehne2024}. 
This connects our work to 
the findings of Eggeling and Werner~\cite{EggelingWerner2001}, 
characterizing the $U^{\otimes 3}$-invariant state space of three
particles. Note, that this kind of symmetry can simplify the solution to hard problems in entanglement theory \cite{EggelingWerner2001, HuberKlepMagronVolcic2022} and is relevant for applications \cite{Cabello2003, Gaertner2008, HilleryBuzek2001}. In fact, our results will later lead to a direct solution
of the problem of GME in
$U^{\otimes 3}$-invariant states.

Moreover, the (anti)chiral subspaces fulfill this $U^{\otimes 3}$-invariance as well. This
shows in different ways. First, it follows from
the fact that $\Pi_\mathcal{J}$ and $\Pi_\mathcal{\Bar{J}}$ are linear combinations of permutations and standard results of the
Schur-Weyl duality~\cite{Weyl1946, EggelingWerner2001}.
Second, by direct computation, one finds
that the ground space of a $U^{\otimes 3}$-invariant observable is also invariant.  The latter argument will 
become important later, as it also holds if one has only
a symmetry with respect to some subgroup of the unitaries.

%%%%%%%%%%%%%%%%%%%%%%%%%%%%%%%%%%%%%%%%%%%%%%%%%%
\textit{High-dimensional generalizations.---}
%%%%%%%%%%%%%%%%%%%%%%%%%%%%%%%%%%%%%%%%%%%%%%%%
Let us generalize the previous results to larger local dimensions $d\geq 3$, 
where we can identify even higher entangled spaces. Now 
there is an antisymmetric subspace as well 
and we have 
$\mathcal{H}=(\mathbb{C}^d)^{\otimes 3} 
= 
\mathcal{H}_{\mathcal{S}}^d \oplus \mathcal{H}_{\mathcal{A}}^d \oplus 
\mathcal{H}_{\mathcal{J}}^d \oplus \mathcal{H}_{\Bar{\mathcal{J}}}^d$. 
The projectors onto the symmetric and antisymmetric subspace read
\begin{align}
    \Pi_\mathcal{S} &= \frac{1}{6}(\mathds{1}+F_{12}+F_{23}+F_{13}+T+T^2),
    \nonumber
    \\
    \Pi_\mathcal{A} &= \frac{1}{6}(\mathds{1}-F_{12}-F_{23}-F_{13}+T+T^2),
\end{align}
where $F_{ij}$ is the flip operator acting on systems $i$ and $j$ 
and $T$ and $T^2$ are the cyclic and anticyclic translation operators 
introduced above, but now acting on local dimension $d$. Taking the sum 
of the projectors $\Pi_{\mathcal{S}\oplus\mathcal{A}}=\Pi_\mathcal{S}+\Pi_\mathcal{A}$, the flip operators cancel out and $\Pi_{\mathcal{S}\oplus\mathcal{A}}$ has the same form as the projector $\Pi_\mathcal{S}$ in two dimensions [see Eq.~(\ref{eq-pisindequal2})]. Consequently, also the projectors on the high-dimensional (anti)chiral subspaces are of the form given in Eq.~(\ref{eq:chiralprojector}). 
Hence, in order to identify highly-entangled spaces we can proceed 
analogously to the qubit case and find:

\textbf{Observation 2.} \textit{For higher-dimensional systems 
($d \geq 3$),  any pure state in the chiral space 
$\mathcal{H}_{\mathcal{J}}$  (or, equivalently, the antichiral 
space $\mathcal{H_{\Bar{\mathcal{J}}}}$) has a geometric measure
of at least $G(\ket{\psi}) \geq \nicefrac{5}{9}$. All states in the 
direct sum $\mathcal{H}_{\mathcal{J}}\oplus \mathcal{H}_{\Bar{\mathcal{J}}}$ 
have an entanglement of at least $G(\ket{\psi})\geq\nicefrac{1}{4}.$}

The proof is a direct generalization of the 
arguments leading to Observation~1, see Appendix~C for details.

Interestingly, within these spaces even higher entangled subspaces can
be identified, so we consider the chiral subspace in more detail. 
Its basis vectors are of the form
\begin{align}
    \ket{\varphi_\alpha} &= \frac{1}{\sqrt{3}}(\ket{iij}+\omega\ket{iji}+\omega^2\ket{jii}),
    \nonumber 
    \\
    \ket{\varphi_{\beta}} &= \frac{1}{\sqrt{3}}(\ket{jji}+\omega\ket{jij}+\omega^2\ket{ijj}),
     \nonumber
     \\
    \ket{\varphi_{\gamma}} &= \frac{1}{\sqrt{3}}(\ket{ijk}+\omega\ket{jki}+\omega^2\ket{kij}),
     \nonumber
    \\
    \ket{\varphi_{\delta}} &= \frac{1}{\sqrt{3}}(\ket{ikj}+\omega\ket{kji}+\omega^2\ket{jik}),
\end{align}
where $i<j<k \in \{0,...,d-1\}$. Note that there are $\binom{d}{2}$ 
basis vectors of types $\ket{\varphi_{\alpha}}$ and $\ket{\varphi_{\beta}}$ 
and $\binom{d}{3}$ basis vectors of types $\ket{\varphi_{\gamma}}$ and 
$\ket{\varphi_{\delta}}$. Therefore, the total dimension of the chiral subspace 
is $\dim(\mathcal{H}_{\mathcal J}^d) = 2\binom{d}{2}+2\binom{d}{3} = {d(d^2-1)}/{3} \sim \mathcal{O}(d^3)$.

So, the chiral subspace splits into two further subspaces, 
$\J = \Jone \oplus \Jtwo$, where $\Jone$ is spanned by the 
$2\binom{d}{2}$ vectors of types $\ket{\varphi_\alpha}$ and 
$\ket{\varphi_{\beta}}$ and $\Jtwo$ by the $2\binom{d}{3}$ 
vectors of types $\ket{\varphi_{\gamma}}$ and $\ket{\varphi_{\delta}}$, 
respectively. Note that, in contrast to the full chiral subspace, 
$\Jone$ and $\Jtwo$ are not $U^{\otimes 3}$-invariant. 
For local dimension $d=3$, however, the subspace 
$\Jtwo$ has interesting properties.

First, for qutrits $\Jtwo$ is a two-dimensional subspace, spanned 
by $\ket{\varphi_{\gamma=(0,1,2)}}$ and $\ket{\varphi_{\delta=(0,1,2)}}$ with indices
$(i,j,k) = (0,1,2)$. These two states are \textit{absolutely maximally entangled} (AME), meaning that all 
reductions of the state to one party (in general, to $\lfloor 
{\#\text{parties}}/{2}\rfloor$) are maximally mixed~\cite{HuberGuehneSiewert2017, Scott2004, RatherBurchardtBruzdaRajchelMieldziocLakshminarayanZyczkowski2022}.
Second, we write all states in this subspace in the Schmidt decomposition in bipartition $A|BC$,
$\ket{\psi} = \nicefrac{1}{\sqrt{3}}\sum_i\ket{i}\ket{\chi_i}$, and from the definition of 
$\ket{\varphi_{\gamma}}$ and $\ket{\varphi_{\delta}}$ follows that the $\ket{\chi_i}$ are normalized.
Then, when maximizing the overlap with product states $\ket{abc}$, we find $|\braket{abc}{\psi}|^2 \leq\nicefrac{1}{3}$  
yielding an entanglement of $G(\ket{\psi})\geq \nicefrac{2}{3}$.

This has two implications. First, $\Jtwo$ is quantitatively more 
entangled than $\J$. Second, since all states in $\Jtwo$ can 
be written in a Schmidt decomposition as above it follows~\cite{GuehneToth2009}
that the largest overlap with biseparable states is also bounded by 
$|\bra{\eta_A}\bra{\mu_{BC}}\psi\rangle|^2 \leq\nicefrac{1}{3}$,
coinciding with the bound for fully separable states. One can generalize the notion of the geometric measure by maximizing the overlap over biseparable states instead of product states~\cite{SenDeSen2010, PrabhuPatiSenDeSen2012}.
Hence, $\Jtwo$ 
is a genuine entangled subspace~\cite{DemianowiczAugusiak2018, DemianowiczAugusiak2019},
where all states have a generalized geometric measure of
entanglement
of $G_{\rm GME}(\ket{\psi}) = \nicefrac{2}{3}$. This is indeed the maximally
possible value, as for pure states in $\mathbb{C}^3 \otimes \mathbb{C}^D$, with $D \geq 3$, the largest Schmidt coefficient
cannot be smaller than $\nicefrac{1}{\sqrt{3}}$. In this way, $\Jtwo$ has analogous 
properties to the space $\mathcal{H}_{\mathcal{J}}$ for three qubits in Observation 1. We summarize:

\textbf{Observation 3.}
\textit{The chiral symmetric three-qutrit subspace $\Jtwo$, which is spanned by two 
AME vectors $\ket{\varphi_{\gamma = (0,1,2)}}$ and $\ket{\varphi_{\delta = (0,1,2)}}$ does not contain any biseparable 
or fully separable states. All states therein have a generalized geometric measure 
$G_{\rm GME}(\ket{\psi}) = \nicefrac{2}{3}$. 
This is the maximal possible value for three-qutrit spaces.}

Finally, note that concerning the standard geometric measure, $G(\ket{\psi}) \geq \nicefrac{2}{3}$
is only a lower bound and some states in $\Jtwo$ are even higher entangled.

\textit{Applications: Entanglement detection.---}
We start by connecting the chiral subspaces to witness operators 
similar to the qubit case. We focus on generalizing the local dimension here, while in Appendix~B we discuss the $N$-particle case. Interpreting the Levi-Civita symbol in Eq.~(\ref{eq:XA2}) as structure constants of the Lie-algebra 
$\mathit{su}(2)$ with Pauli matrices as basis operators via $[\sigma_i,\sigma_j]=2\textbf{i}\sum_k\varepsilon_{ijk}\sigma_k$,  a natural generalization to higher dimensions uses Gell-Mann matrices and the corresponding 
structure constants of the Lie-algebra $\mathit{su}(d)$, 
\begin{align}
    \kappa_{ijk}^- = -\frac{\textbf{i}}{d^2}\tr(\lambda_i[\lambda_j,\lambda_k]), 
    \;\;
    \kappa_{ijk}^+ = \frac{1}{d^2}\tr(\lambda_i\{\lambda_j,\lambda_k\}),
\end{align}
with commutator $[A,B] = AB-BA$ and  anticommutator $\{A,B\}=AB+BA$. The coefficients $\kappa_{ijk}^-$ are the natural generalization of the Levi-Civita Symbol while $\kappa_{ijk}^+$ is nontrivial only for $d\geq 3$. The operators $\lambda_i$ are the generalized traceless Gell-Mann matrices in $d$ dimensions and fulfill the norm $\tr(\lambda_i\lambda_j)=d\delta_{ij}$~\cite{Georgi1982}. 
Replacing $\sigma_i\mapsto \lambda_i$ and $\epsilon_{ijk}\mapsto \kappa^\pm_{ijk}$ we define two observables,
\begin{align}
W_\pm = \sum_{i,j,k=1}^{d^2-1}\kappa^\pm_{ijk}\lambda_i\otimes\lambda_j\otimes\lambda_k,\label{eq:Wminus}
\end{align} 
where $W_-$ is the natural generalization of $W_\varepsilon$ in Eq.~(\ref{eq:XA2}) and $W_+$ only exists for local dimension $d\geq3$.
These observables are useful for entanglement detection in the following sense:

\textbf{Observation 4.} \textit{The observables $W_\pm$ can be used as entanglement witnesses. Their expectation values for fully separable 
(fs), biseparable (bs) and general quantum states (q) are bounded by 
\begin{align}
    \langle W_-\rangle &\,\stackrel{\rm fs}{\leq}\,
    \frac{d}{2} \, \overset{\rm bs}{\leq}\, 
    d \, \overset{{\rm q}}{\leq}\, d\sqrt{3}
    \;\;{\rm for }\;\; d \geq 3,
    \\
    \langle W_+\rangle &\,\stackrel{\rm fs}{\leq}\,
    \frac{4}{3}\, \overset{\rm bs}{\leq}\,
    \frac{10}{3}\, \overset{{\rm q}}{\leq}\, \frac{40}{3} \;\;{\rm for }\;\;d=3, 
    \nonumber
    \\
    \langle W_+\rangle &  \, \overset{{\rm fs\&bs}}{\leq}\, 
    \frac{2(d-1)(d-2)}{d}\, \overset{{\rm q}}{\leq}\,
    \frac{2(d+1)(d+2)}{d} \;\;{\rm for }\;\;d\geq 4.
    \nonumber 
\end{align}
}
Detailed proofs are given in Appendices~E and~F; they rely on 
several properties of the observables $W_\pm$, which are of 
independent interest and deserve some discussion.

First, the definition of $W_\pm$ is independent of the operators $\lambda_i$, i.e., replacing these 
by any set of hermitian, orthogonal and traceless operators 
$\{G_i\}$ fulfilling $\tr(G_iG_j)=d\delta_{ij}$ yields
identical operators $W_\pm$.

Second, as the qubit operator in Eq.~(\ref{eq:XA2}) the 
observables $W_\pm$ are invariant under $U^{\otimes 3}$-transformations
and thus are decomposable into linear combinations 
of permutations~\cite{Weyl1946, EggelingWerner2001}. This is useful for showing the separability bound.

Third, considering the operators' spectrum, we find 
as in the three-qubit case that $W_- = \alpha(\Pi_{\Bar{\mathcal{J}}}-\Pi_\mathcal{J})$ with  $\alpha = d\sqrt{3}$. The operator $W_+$ has 
full rank and can be written as $W_+ = c_\mathcal{S}\Pi_\mathcal{S} +c_\mathcal{A}\Pi_\mathcal{A} + c_\mathcal{J}(\Pi_\mathcal{J}+\Pi_{\bar{\mathcal{J}}})$, with $c_\mathcal{S} = \nicefrac{2}{d}(d-1)(d-2)$, $ c_\mathcal{A} = \nicefrac{2}{d}(d+1)(d+2)$ and $c_{\mathcal{J}} = -\nicefrac{1}{d}(d+2)(d-2)$.
The largest eigenvalue corresponds to the antisymmetric basis vectors and 
implies the quantum bound for $W_+$. 
Note that from this representation
also follows that $W_-$ and $W_+$ commute.

Finally, for the biseparability bounds we can make use of the $U^{\otimes 3}$-invariance and assume that the bound is reached by a state 
$\ket{\eta_A}\ket{\mu_{BC}}=\ket{0}\ket{\mu_{BC}}$, boiling down the optimization to finding the largest eigenvalue of the conditional observable 
$\bra{0} W_\pm \ket{0}$ on the system $BC$. 
This, indeed, has far-ranging consequences beyond this example.
Assume, one wishes to decide GME of a given $U^{\otimes 3}$-invariant state $\varrho$. Then, it suffices to consider 
entanglement witnesses $\mathcal{W}$ with the same symmetry, and 
such operators are linear combinations of six 
permutation operators~\cite{EggelingWerner2001}. Effectively, 
one only needs to minimize $\tr(\varrho \mathcal{W})$ under the 
constraint that the conditional observables $\bra{0} \mathcal{W} 
\ket{0}$ have no negative eigenvalues for the systems $AB$, $BC$, and 
$AC$, solving the separability problem in this case by a 
simple semidefinite programme.

\begin{figure}[htbp]
    \centering
    \includegraphics[width=0.9\linewidth]{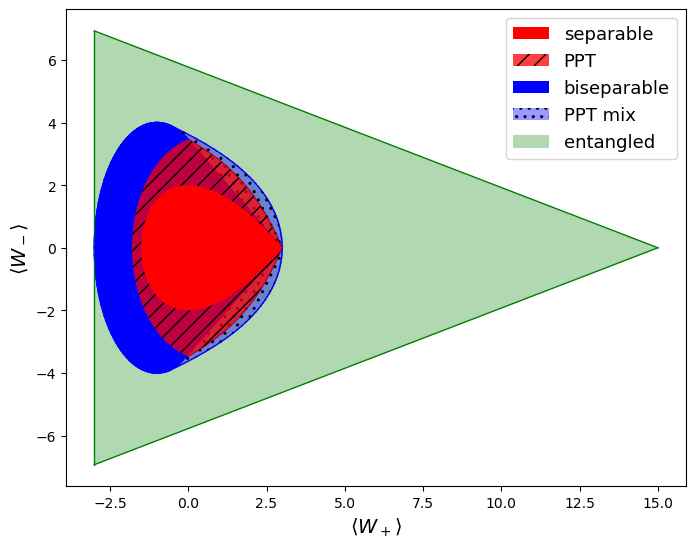}
    \caption{State space of three ququads parametrized by the expectation values of the operators $W_\pm$. We differentiate between states that are fully separable (red), PPT for all bipartitions (red hatched), biseparable (dark blue), a mixture of states that are PPT for one bipartition (red, dark blue, and blue dotted) and GME (blue dotted and green). Since the operators $W_\pm$ commute, the state space forms a polytope. The three vertices of the triangle contain the states of the chiral, antichiral and antisymmetric subspace, respectively. 
    We further find that combining the two witnesses allows the detection of PPT entangled states. In fact, the red hatched area intersecting with the blue dotted one gives rise to PPT states that are GME. The state space looks similar for qutrits, see also Appendix~G. Moreover, a similar Figure for more general witnesses is given in Appendix~I}
    \label{fig:witness}
\end{figure}

Furthermore, other combinations of the observables 
$W_\pm$ yield strong entanglement witnesses, see  Fig.~\ref{fig:witness}. These can detect 
states as GME that have a positive partial transpose (PPT) 
for any bipartition and, therefore, cannot be detected 
by the commonly 
used PPT criterion~\cite{HorodeckiHorodeckiHorodecki1996, Peres1996}. 
For three qubits, the 
existence of such states with the given symmetry was ruled out~\cite{EggelingWerner2001}, and in higher dimensions our states
are the first examples. 
We summarize: 

\textbf{Observation 5.}
\textit{(a) A single semidefinite programme can decide
the question of GME for 
a given tripartite $U^{\otimes 3}$-invariant state.
(b) The observables
\begin{align}
    P = \Pi_\mathcal{J}-\Pi_\mathcal{A}\,\,\text{and}\,\, \bar{P}=\Pi_{\bar{\mathcal{J}}}-\Pi_\mathcal{A},
\end{align}
are  witnesses 
for GME. They detect GME 
states which are PPT for all bipartitions.
}

The proof relies on the same methods described above and is given in full detail in Appendix~H, where we also characterize the detected PPT states.

%%%%%%%%%%%%%%%%%%%%%%%%%%%%%%%%%%%%%%%%%%%
\textit{Conclusion.---}
%%%%%%%%%%%%%%%%%%%%%%%%%%%%%%%%%%%%%%%%%%%
In conclusion, we have shown that chiral symmetry and 
multiparticle entanglement are inextricably interwoven.
Chiral symmetries allow us to identify highly-entangled 
subspaces, which have various connections to entanglement 
detection, generalized measurements, and may find 
applications in data hiding~\cite{HayashiMarkhamMuraoOwariVirmani2006, TerhalDiVincenzoLeung2001}. 
We identified 
$U^{\otimes 3}$-invariant GME states which are PPT for any
bipartition and showed that GME for $U^{\otimes 3}$-invariant
states can be characterized in a straightforward manner. 
In Appendix~A, we further consider a relaxed symmetry, namely of states that 
become complex conjugated when exchanging two particles: $F_{ij}\ket{\psi} = \ket{\psi^*}$.

Our results open the door to several routes of research that may be pursued in the future.
 
First, it would be relevant to understand the geometric measure for states with certain symmetries.
For instance, it is well known that the closest 
product state to a fully symmetric state needs to be symmetric, 
so it is relevant to formulate similar statements for the other 
spaces. This could also help to understand how to identify the maximally entangled states with respect to the GM in general.
Second, from an experimental perspective, it would be 
very useful to formulate robust statements, giving lower bounds on the entanglement if 
a state is approximately in some symmetric subspace. Third, our work seems to be strongly connected to Bargmann invariants, so exploring this, also in the context of Ref.~\cite{Wagnerneu} would be desirable.
Finally, while the geometric measure has been used beyond entanglement,
e.g., stabilizer rank~\cite{BravyiSmithSmolin2016}, matrix product states~\cite{Nico-KatzBose2023}, or error-correcting codes
\cite{BravyiLeeLiYoshida2025, LiLeeYoshida2025}
one may also investigate these possibilities for chiral symmetries.

{\it Note added.---} While finishing this work we learned about a related 
work~\cite{RicoZawKrebs2025} which constructs entanglement witnesses for 
$U^{\otimes 3}$-invariant states in a complementary manner. Interestingly, that work implies that some of our witnesses can be simulated efficiently on a quantum computer as well and moreover give additional new immanant inequalities.

%TC:ignore

\textit{Acknowledgements.---}
We thank Adam Burchardt,
Serge Deside, Ernesto Galv\~ao, Kenneth Goodenough, Tobias Haas,
Felix Huber, Johannes Knörzer, Owidiusz Makuta,
Carlo Marconi, Ties Ohst, Albert Rico, Konrad Szymanski, Ryuji Takagi, Julio de Vicente, 
Giuseppe Vitagliano, Rafael Wagner, Lisa Weinbrenner, and Nikolai Wyderka
for discussions. This work has been supported 
by the Deutsche Forschungsgemeinschaft (DFG, German Research
Foundation, project numbers 447948357, 440958198 and 563437167),
the Sino-German Center for Research Promotion (Project 
M-0294), the German Ministry of Education and Research 
(Project QuKuK, BMBF Grant No. 16KIS1618K), and the German Federal Ministry of Research, Technology
and Space (Projects BeRyQC, Grant No. 01N124023D, and PraQTisch, Grant No. 01N126223D).
S.D. acknowledges support by the House of Young Talents of the University of Siegen.
S.I. acknowledges support by Horizon Europe programme HORIZON-CL4-2022-QUANTUM-02-SGA via the project 101113690 (PASQuanS2.1) and JST ASPIRE (JPMJAP2339).

\onecolumngrid
\appendix

\section{The flip-conjugate-symmetric subspace}
We now want to discuss a relaxed symmetry, called \textit{flip-conjugate symmetry}. Precisely, we consider states that become complex conjugated when exchanging two particles: $F_{ij}\ket{\psi} = \ket{\psi^*}$. Depending on which of the three particles remains untouched, these states define three different, but analogously behaving subspaces and their complex conjugate. In the following we focus on the one acting under $F_{23}$, denoted $\mathcal{H_I}$. This subspace has some intriguing properties.

First, the basis vectors are eigenvectors of the partially transposed observable $W_-$ in Eq.~(\ref{eq:Wminus}). Consequently, $W_-^{T_1}$ and thus also the subspace $\mathcal{H_I}$ are $U^*\otimes U\otimes U$-invariant~\cite{Weyl1946, EggelingWerner2001}, see Appendix~I for more details.

%%%%%%%%%%%%%%%%%%%%%%%%%%%%%%%%%%%%%%%%%%%%%

We can now use the partially transposed operator $\mathbb{W}_- = W_-^{T_1}$ to identify the basis vectors of the subspace $\mathcal{H_I}$.
In fact, it has two 
nonzero eigenvalues which are $d$-fold degenerate each. The 
corresponding eigenvectors are given by
\begin{align}
    \ket{\phi_n} &= \frac{1}{\sqrt{d+1}}
    \left (z_d^*\sum_{i=0}^{d-1}\ket{ini}+z_d\sum_{i=0}^{d-1}\ket{iin} \right)\,\,\text{and}\label{eq:basisUsUU}\\
    \ket{\bar{\phi}_n} &= \frac{1}{\sqrt{d+1}}
    \left(z_d\sum_{i=0}^{d-1}\ket{ini}+z_d^*\sum_{i=0}^{d-1}\ket{iin} \right),
\label{eq:basisUsUUbar}
\end{align}
for $n = (0,...,d-1)$, with $z_d=\nicefrac{1}{2}(1+\mathbf{i}\sqrt{({d+1})/({d-1})})$. 
Hence, the projectors onto $\mathcal{H_I}$ and its complex conjugate $\mathcal{H}_{\bar{\mathcal{I}}}$ read
\begin{align}
\Pi_\mathcal{I} &= \frac{1}{d+1} \left[|z_d|^2(\mathbb{F}_{12}+\mathbb{F}_{13})+(z^*_d)^2\mathbb{T}^2+z_d^2\mathbb{T} \right],\\
\Pi_{\bar{\mathcal{I}}} &= \frac{1}{d+1} \left[|z_d|^2(\mathbb{F}_{12}+\mathbb{F}_{13})+z_d^2\mathbb{T}^2+(z_d^*)^2\mathbb{T} \right],
\end{align}
where $\mathbb{T} = T^{T_1}$ and $\mathbb{F}_{ij} = F_{ij}^{T_1}$ are the partially transposed translation and flip operators.

Second, all states in $\mathcal{H_I}$ are local unitary (LU) equivalent:

\noindent
\textbf{Proposition 6.} \textit{All states in $\mathcal{H_I}$ (and analogously in $\mathcal{H}_{\bar{\mathcal{I}}}$) are LU equivalent and can be reached by applying a local unitary of the form $U^*\otimes U \otimes U$ to $\ket{\phi_0}$.}
\begin{proof}
    Without loss of generality, we take the vector $\sum_i\ket{ii0}$ and consider:
    \begin{align}
        &U^*\otimes U \otimes U \sum_i\ket{ii0} = (U^*\otimes U )\ket{\phi^+}\otimes U \ket{0} \notag
        \\= &\ket{\phi^+} \otimes \sum_n\alpha_n\ket{n} = \sum_{ni}\alpha_n\ket{iin},
    \end{align}
    where we take $\ket{\phi^+} = \sum_i\ket{ii}$ as the unnormailzed maximally entangled state and use that it is invariant under $U^*\otimes U$-transformations. Further, applying a unitary to $\ket{0}$ gives an arbitrary linear combination of basis vectors in the computational basis, i. e. $U\ket{0}=\sum_n\alpha_n\ket{n}$, were the coefficients $\alpha_n$ are the entries of the first column of $U$. 

    Analogously, it holds $U^*\otimes U \otimes U \sum_i\ket{i0i} = \sum_{ni}\alpha_n\ket{ini}$ and consequently, an arbitrary superposition $\ket{\psi} = \sum_n \alpha_n \ket{\phi_n}$ with normalized coefficients $\alpha_n$ can always be written as $U^*\otimes U \otimes U\ket{\phi_0}$, if we choose $U$ such that it has the 
    $\alpha_n$ as entries in the first column.
\end{proof}
The LU equivalence is also useful to show that all states in the flip-conjugate subspace are highly entangled.

\noindent
\textbf{Observation 7.}
\textit{For systems in dimension $d \leq 12$,  any pure state in the 
$d$-dimensional {flip-conjugate symmetric} space 
$\mathcal{H}_{\mathcal{I}}$  (or, equivalently, 
the space $\mathcal{H_{\Bar{\mathcal{I}}}}$) has 
a geometric measure $G(\ket{\psi}) = 1-{d^2}/[{(d+1)(d^2-1)}]$.
For arbitrary dimensions, the bound 
$G(\ket{\psi}) \geq 1-{d}/({d^2-1})$ holds.
}
\begin{proof}
For small local dimensions we can derive a bound using a semidefinite programme (SDP). We consider $\max_{\ket{abc}}
\bra{abc}\Pi_\mathcal{I}\ket{abc}$ and, due to the 
$U^*\otimes U\otimes U$-invariance, we can fix 
$\ket{a}=\ket{0}$ and compute a relaxation~\cite{WeinbrennerGuehne2025}
$\max_{\ket{bc}}\bra{bc}\bra{a}\Pi_\mathcal{I}\ket{a}\ket{bc}\leq \max_{\varrho^\text{PPT}}\tr(\bra{a}\Pi_{\mathcal{I}}\ket{a}\varrho^\text{PPT})$. This is an SDP and yields $G(\ket{\psi}) = 1-{d^2}/[(d+1)(d^2-1)]$.

For higher dimensions, an SDP becomes impractical, but making use of the fact that all states in the flip-conjugate symmetric subspace are LU equivalent, we can give a bound on the geometric measure by maximizing the overlap for one basis vector $\ket{\phi_0} = \nicefrac{1}{\sqrt{d+1}}(z_d^*\sum_{i=0}^{d-1}\ket{i0i}+z_d\sum_{i=0}^{d-1}\ket{ii0})$.

Note that we can write $\ket{\phi_0} = \nicefrac{\sqrt{d}}{\sqrt{d+1}}(z_d^*F_{23}\ket{\gamma}+z_d\ket{\gamma},
$
where $\ket{\gamma} = \nicefrac{1}{\sqrt{d}}\sum_{i=0}^{d-1}\ket{ii0}$. Then, we can compute the overlap
\begin{align}
    \braket{abc}{\phi_n} &= \sqrt{\frac{d}{d+1}}(z_d^*\braket{acb}{\gamma}+z_d\braket{abc}{\gamma})\notag\\
    &= \sqrt{\frac{d}{d+1}}(\bra{a}\otimes ( z_d^*\bra{cb} + z_d\bra{bc})\ket{\gamma}\notag \\&=
    \sqrt{\frac{d}{d+1}}\sqrt{\mathcal{N}}\langle\varphi^{A|BC}|\gamma\rangle,
\end{align}
where $|\varphi^{A|BC}\rangle = \ket{a}\otimes(\tilde{z}_d\ket{cb}+\tilde{z}_d^*\ket{bc})$ with $\tilde{z}_d = \nicefrac{z_d}{\sqrt{\mathcal{N}}}$ and $\tilde{z}^*_d = \nicefrac{z^*_d}{\sqrt{\mathcal{N}}}$ is a normalised state, which is biseparable with respect to the bipartition $A|BC$ and $\mathcal{N} = 2|z_d|^2+2\text{Re}[(z_d^*)^2]|\braket{c}{b}|^2$ its norm. 

We can now use the fact that for a given $\ket{\psi}$ the overlap $\max_{|\varphi^{\text{bisep}\,A|BC}\rangle}|\langle\varphi^{\text{bisep}\,A|BC}|\psi\rangle| = s^{A|BC}_1$ 
 is maximized by the largest Schmidt coefficient of $\ket{\psi}$ in the bipartition $A|BC$~\cite{SanperaBrussLewenstein2001}.
Then, considering the Schmidt decomposition $\ket{\gamma} = \sum_i\nicefrac{1}{\sqrt{d}} \ket{i}\otimes\ket{i0}$ in the bipartition $A|BC$, we find that $s_1^{A|BC}=\nicefrac{1}{\sqrt{d}}$. This yields
\begin{align}
    &\max_{\ket{abc}}|\braket{abc}{\phi_n}| = \max_{\ket{abc}}\sqrt{\frac{d}{d+1}}\sqrt{\mathcal{N}}\left|\langle\varphi^{A|BC}|\gamma\rangle\right|\notag \\
    &\leq \max_{\ket{\varphi^{\text{bisep}\,\,A|BC}}}\sqrt{\frac{d}{d+1}}\sqrt{\mathcal{N}}\langle\varphi^{A|BC}|\gamma\rangle = \sqrt{\frac{\mathcal{N}}{d+1}}\leq \sqrt{\frac{d}{d^2-1}},
\end{align}
where in the last step we use that $\mathcal{N}\leq 2|z_d|^2$, since $\text{Re}[(z_d^*)^2]<0$ for all $d$.
\end{proof}
Note that the proof for higher dimensions gives only an upper bound on the overlap, since in the first inequality we replace the maximization over all separable states by a maximization over all biseparable states with respect to partition $A|BC$ instead of considering only biseparable states of the specific form $|\varphi^{A|BC}\rangle = \ket{a}\otimes(\tilde{z}_d\ket{cb}+\tilde{z}_d^*\ket{bc})$.

\begin{figure}[t!]
    \centering
    \includegraphics[width=0.6\linewidth]{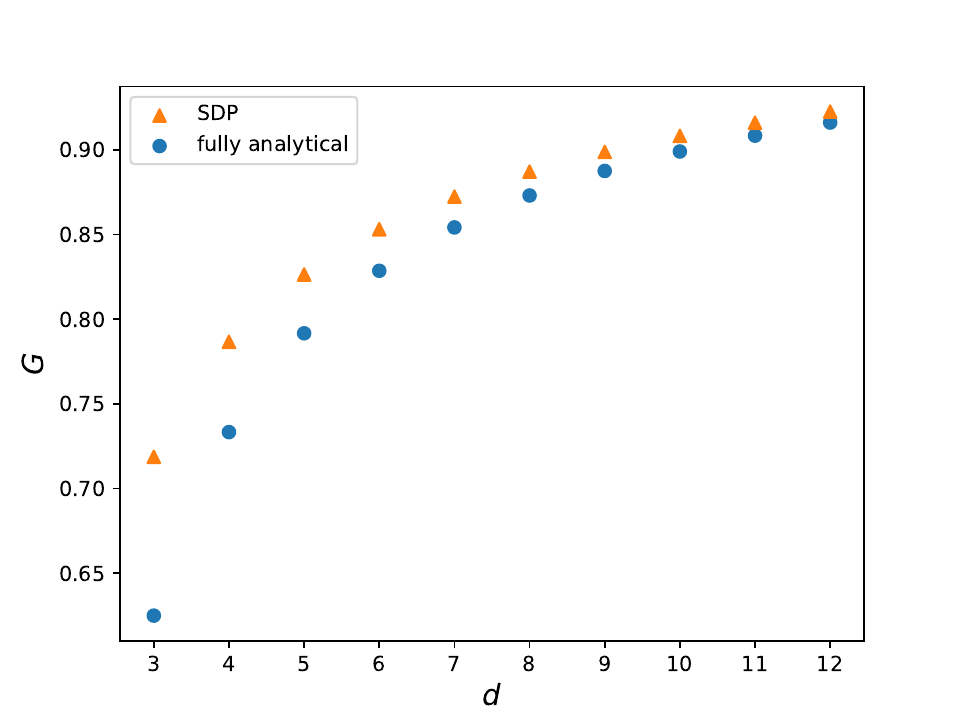}
    \caption{Comparison of the results for the geometric measure: The SDP results are tighter bounds, but cannot be obtained for arbitrary large dimensions, the analytical values hold for all dimensions $d$.}
    \label{fig:comparisonGeom}
\end{figure}

In Fig.~\ref{fig:comparisonGeom} we compare the results obtained by the SDP with the fully analytical method. Note that while the SDP gives a tighter bound, it cannot be solved for arbitrary large dimensions $d$. On the other hand, the fully analytical bound is valid for all dimensions and we can see that for dimension $d=10$ it already gives a very good approximation to the result obtained by the SDP. We remark that the $d=2$ case is excluded in this plot, since for qubits we can find unitaries such that $U_A\otimes U_B\otimes U_C \Pi_\mathcal{J} U_A^\dagger\otimes U_B^\dagger\otimes U_C^\dagger  = \Pi_{\mathcal{I}}$ meaning that the flip-conjugate symmetric subspace is LU equivalent to the chiral symmetric subspace and has therefore geometric measure equal to $\nicefrac{5}{9}$.

Moreover, we observe that the geometric measure of the flip-conjugate symmetric subspace strongly increases with the local dimension $d$: $G(\ket{\psi}) \sim 1-\mathcal{O}(\nicefrac{1}{d})$, yielding $G(\ket{\psi})\rightarrow 1$ for large $d$. This implies its usefulness for encoding quantum information with built-in secret sharing, since it is known that states, which are encoded in a highly entangled subspace are hard to discriminate via local operations and classical communication (LOCC), which is useful for applications such as quantum data hiding~\cite{HayashiMarkhamMuraoOwariVirmani2006, TerhalDiVincenzoLeung2001}.

Lastly, the strucute of the flip-conjugate symmetric states inspires us to consider generalizations of LU equivalent highly entangled subspaces. First, note that the fact that $\mathcal{H_I}$ is LU equivalent does not depend on the choice of the complex coefficients $z_d$ in the basis vectors. So, we can replace $z_d'
=e^{\frac{\pi}{2}\mathbf{i}} = \mathbf{i}$ or equivalently $z_d'= e^{\frac{3\pi}{2}\mathbf{i}} = -\mathbf{i}$ in Eq.~(\ref{eq:basisUsUU}) (or Eq.~(\ref{eq:basisUsUUbar})), which yields a subspace that is antisymmetric with respect to the flip of Bob's and Charlie's system. This subspace has geometric measure $G(\ket{\psi}) = 1-\nicefrac{1}{[2(d-1)]}$ for dimensions $d=3,4,5$, which are values close to the largest possible values~\cite{SteinbergGuehne2024}.

Note that this gives rise to further generalizations. In fact we can always take a highly entangled $U^*\otimes U^{\otimes N-2}$-invariant $N-1$-particle state and add a party $\ket{n}$, then summing over permutations and choosing the coefficients $z_d$ accordingly, we can construct $U^*\otimes U^{\otimes N-1}$-invariant highly entangled $N$-particle subspaces. In particular, also the chiral symmetric subspace can be interpreted and generalized accordingly, which we will see in the following.

Let us start considering the 
three-qubit case. Here, we know that the subspace is $U^{\otimes 3}$-invariant. Moreover, in the qubit case all states are LU equivalent and can be reached by $U^{\otimes 3} \ket{\phi_0}$~\cite{DerksenFriedlandLimWang2017, LiuKnoerzerWangTura2024, AlbertiniDAlessandro2021}. As above, we consider the state
\begin{align}
    \ket{\psi_0} = \mathcal{N} \left( x \ket{\phi^-}_{AB}\ket{0}_C + y \ket{\phi^-}_{AC}\ket{0}_B + z \ket{\phi^-}_{BC}\ket{0}_A\right),
\end{align}
where $\ket{\phi^-} = \ket{01}-\ket{10}$ is the unnormalized $U\otimes U$-invariant Bell-State. Choosing the parameters 
$x,y$ and $z$ such that 
\begin{align}
    x-z = \omega, \quad
    -x-y = \omega^2
  , \quad
    y+z = 1,
\end{align}
we can reconstruct the chiral symmetric three-qubit state 
$\ket{\phi_1}$ and correspondingly the entire space 
$\mathcal{H_J}.$ 

One can generalize this approach to more parties by taking the highly entangled $U^{\otimes 3}$-invariant fully antisymmetric three-qutrit state $\ket{\psi^-_3}$:
\begin{align}
    \ket{\psi_0} = \mathcal{N}(x\ket{\psi^-_3}_{ABC}\ket{0}_D 
    + y\ket{\psi^-_3}_{ABD}\ket{0}_C\\+z\ket{\psi^-_3}_{ACD}\ket{0}_B+k\ket{\psi^-_3}_{BCD}\ket{0}_A ).
\end{align}
Then, choosing the phases $x,y,z$ and $k$ accordingly, for example such that 
\begin{align}
    x =1, \quad
    y = \omega_4^3, \quad
    z = \omega_4^2, \quad
    k = \omega_4
\end{align}
with $\omega_4 = e^{\mathbf{i}\frac{\pi}{2}} = \mathbf{i}$, yields a state that is chiral symmetric, (i.e., $T\ket{\psi} = \mathbf{i} \ket{\psi}$) and has geometric measure $G(\ket{\psi_0})=\nicefrac{7}{8}$. For four qutrits the largest possible value is $G(\ket{\psi_\mathrm{max}}) = \nicefrac{8}{9}$~\cite{SteinbergGuehne2024}. Hence, the state $\ket{\psi_0}$ is highly entangled, with geometric measure close to its maximum. Moreover, by construction, it defines a three-dimensional $U^{\otimes 4}$-invariant subspace, where all vectors are local unitary equivalent. The LU equivalence directly follows from construction and to show the $U^{\otimes 4}$-invariance, consider: 
\begin{align}
    &U^{\otimes 4}\Pi(U^{\otimes 4})^\dagger = \sum_i U^{\otimes 4} \ketbra{\psi_i}(U^{\otimes 4})^\dagger \\\notag
    = &\sum_i\sum_{nm}c_{ni}c_{mi}^*\ket{\psi_n}\bra{\psi_m} = \sum_{nm}\delta_{nm}\ket{\psi_n}\bra{\psi_m} = \sum_n \ketbra{\psi_n},
\end{align}
where in the second step we used that $U\ket{\psi_i} = \sum_n c_{ni}\ket{\psi_n}$ with $c_{ni}$ the coefficients of the $i$th column of $U$ and in the third step that the rows and columns of a unitary are orthogonal: $\sum_{i}c_{ki}c_{li}^*=\delta_{kl}$.

\section{Generalizations to $N$ particles}
In this appendix we want to briefly discuss approaches to generalize our findings to more particles. We will give three different ansatzes and summarize the results in Table~\ref{tab:res}. The first approach generalizes the property of highly entangled subspaces, whereas the second approach focuses on generalizing the aspect of chiral symmetry. The last approach is a generalization of the fact that the chiral subspace in the three-partite case are irreducible representations of the alternating group. 
Lastly, we demonstrate how the entanglement witness $W_-$ arising from the chirality operator, can be generalized to $N$ particles as well.

\subsection{Generalization based on LU equivalent basis vectors}
First, note that one way to generalize the highly entangled subspaces to spaces of $N+1$ particles in local dimension $d = N$ is the one mentioned in Appendix~B. To keep this appendix complete, we briefly recap: The basis vectors are constructed by taking the state $\ket{n}$, where $n \in \{0,\cdots,d-1\}$ on one of the $N+1$ systems and the totally antisymmetric $N$ particle state on the other $N$ parties. Summing over all permutations then yields a basis vector for all $n\in \{0,\cdots,d-1\}$:
\begin{align}
    \ket{\Psi_n} \propto \sum_{i=1}^{N+1}\alpha_i\ket{n}_{A_i}\otimes\ket{\Psi^-_N}_{A_1\cdots A_{N+1}\backslash A_i},
\end{align}
where the coefficients $\alpha_i$ should be chosen such that the geometric measure of $\ket{\Psi_n}$ is as large as possible. For $N+1=4$ a suitable choice is to set either all coefficients equal to one, or take the the 4th root of unity to different powers: $\alpha_1 = \mathbf{i}, \alpha_2 = -1,\alpha_3 = -\mathbf{i}$ and $\alpha_4 =1$.

This construction yields an $N$-dimensional subspace in which all states are local unitary equivalent and the full subspace is invariant under $U^{\otimes N+1}$ transformations. Bounding the geometric measure with a see-saw algorithm from above and with a semidefinite programme from below \cite{WeinbrennerGuehne2025} shows that all states in the four-particle subspace have a geometric measure $E_G=\nicefrac{7}{8}$. We summarize and compare this example to further methods in Table~\ref{tab:res}.

\subsection{Generalization based on the chirality operator}
Another way to generalize chiral subspaces to more particles is by generalizing the chirality operator (Eq.~(7) in the main text) to $N$ particles and building the projectors from that. For three qubits the chirality operator is given by:
\begin{align}
W_\varepsilon \coloneqq \chi_3
= 
\sum_{i,j,k=1}^3\varepsilon_{ijk}\sigma_i\otimes\sigma_j\otimes\sigma_k
= 
2\sqrt{3} (\Pi_{\Bar{\mathcal{J}}}-\Pi_\mathcal{J}) = \mathbf{i}\times 2(T-T^2) \propto \textbf{i}[F_{23},F_{13}],
\end{align}
where the last equation follows the notation in Ref.~\cite{chiralOpN}.  
There, a generalization of the chirality operator is given by
\begin{align}
    \chi_N \propto \textbf{i} [P_1,P_2],
\end{align}
where $P_i$ are permutation operators that can be derived straightforwardly following Ref.~\cite{chiralOpN}. According to Ref.~\cite{chiralOpN}, the chirality operator is known to vanish for even $N$. Hence, the easiest nontrivial generalization is for five qubits. Then we have
\begin{align}
    \chi_5 \propto \mathbf{i}[F_{25}F_{34},F_{13}F_{45}] = \mathbf{i}(T^3-T^2),
\end{align}
where $T$ now acts on five particles.
This operator has a total of 24 nonzero eigenvalues given by $\pm \alpha$ and $\pm \beta$. Hence, we can construct two pairs of chiral and antichiral subspaces, belonging to the eigenvalues $\pm \alpha$ and $\pm \beta$. The projectors read
\begin{align}
    \Pi_{\bar{J}_\alpha} &= \frac{1}{5}(\mathds{1}+\omega_5 T + \omega_5^2 T^2 +\omega^3_5T^3+\omega_5^4T^4)\label{eq:pi1}\\
    \Pi_{J_\alpha} &= \frac{1}{5}(\mathds{1}+\omega^4_5 T + \omega_5^3 T^2 +\omega^2_5T^3+\omega_5T^4),\label{eq:pi2} \\
    \Pi_{\bar{J}_\beta} &= \frac{1}{5}(\mathds{1}+\omega_5^3 T + \omega_5 T^2 +\omega_5^4T^3+\omega_5^2T^4),\\
    \Pi_{J_\beta} &= \frac{1}{5}(\mathds{1}+\omega_5^2 T + \omega_5^4 T^2 +\omega_5T^3+\omega_5^3T^4),
\end{align}
with the cyclic translation operator again acting on five particles and $\omega_5 = e^{\frac{2\pi\mathbf{i}}{5}}$, the 5th root of unity. Note that $\omega_5^* = \omega_5^4$ and $(\omega_5^2)^*=\omega_5^3$ and therefore the chiral subspaces are given by the complex conjugate of the antichiral subspaces as above. Moreover, following the notation for three particles, the chiral subspaces are spanned by the eigenvectors of $\chi_5$ corresponding to the negative eigenvalues, while the antichiral subspaces correspond to the positive eigenvalues. Evaluating an SDP yields $0.5883$ as a lower bound onto the GM for all states in these subspaces (see also Table~\ref{tab:res}). Note however, that since the SDP is just a relaxation method the true lower bound might be larger.

There are two further comments to mention.
First, Eqs.~(\ref{eq:pi1}) and (\ref{eq:pi2}) are the exact generalization of Eq.~(4) in the main text, which are obtained by letting the cyclic translation $T$ act on five particles instead of three and replacing the 3th root of unity by the 5th. However, in the case of five particles there are more possibilities to distribute the phases $\omega_5$ such that $\sum_i\omega_5^{j(i)}T^{k(i)}$ yields a valid projector. This comes from the fact, that while $T$ acting on three qubits has three different eigenvalues $1,\omega_3$ and $\omega_3^2$ defining the symmetric, chiral and antichiral subspace, $T$ acting on five particles has eigenvalues $1,\omega_5,\omega_5^2,\omega_5^3$ and $\omega_5^4$ yielding a subspace of states, fulfilling $T\ket{\psi}=\ket{\psi}$ and two pairs of chiral and antichiral subspaces.

Second, a different way to construct chiral subspaces for more particles is  to diagonalize the cyclic translation operator $T$, which will always have eigenvalues $\omega_N^k$ for $k=0,...,N-1$, and for odd $N$ one eigenvalue will be equal to one and the others are couples of complex phases and their complex conjugate. Then, the eigenvectors corresponding to the complex eigenvalues form the bases of the respective chiral and antichiral subspaces. Note that in fact, since the translation operator is diagonalized by the discrete Fourier transform, the basis vectors will correspond to the Fourier basis.

\subsection{Generalization based on group theory}
An intuitive way to see why all states in the three-qubit chiral symmetric subspace maximize the GM could be the following: Considering the results in Ref.~\cite{SteinbergGuehne2024} one may conjecture that if $N=d$ the geometric measure is always maximized by the totally antisymmetric state. However, if $d<N$ as in the three qubit case, there is no totally antisymmetric subspace, but the \textit{best replacement of the antisymmetric subspace} is connected to the chiral subspace. We will now give a brief explanation why one approach to look for highly entangled states and subspaces in a general way is connected to symmetries. Then we will discuss how we can generalize the chiral subspaces in that sense.

Let us be more specific: According to the Schur-Weyl duality the Hilbert space splits into subspaces of irreducible representations (irreps) that are invariant under $U^{\otimes N}$ transformations as well as under the action of the symmetric group. (For more details on representation theory, see for example \cite{GT1,GT2,GT4}). These irreps can be labeled by partitions $\lambda = [\lambda_1,\cdots ,\lambda_d]$, such that $\lambda_1 \geq \lambda_2\geq \cdots\geq\lambda_d$ and $\lambda_1+\cdots+\lambda_d = N$. In the three qutrit case for example, these are given by $\lambda = [3,0,0],\,[2,1,0]$ and $[1,1,1]$. Then, the first partition corresponds to the symmetric subspace, the last one to the antisymmetric subspace and the second one to some intermediate symmetry. In fact, we have $\mathcal{H} = \mathcal{H}_{\mathcal{S}}\oplus\mathcal{H}_{\mathcal{J}\oplus\bar{\mathcal{J}}} \oplus \mathcal{H_A}$. The lower bound on the GM of the antisymmetric subspace is known to be $1-\nicefrac{1}{N!} = \nicefrac{5}{6}$ \cite{antisymm}. Moreover, for the symmetric subspace, we know that it contains product states, so the lower bound is zero. Finally, for the intermediate symmetry, we will show later (in Appendix C) that the bound on the GM is $\nicefrac{1}{4}$. This suggests some ordering,  stating that the more antisymmetric a subspace is, the higher its geometric measure. 

Before continuing with the generalization of the three-qubit case let us add two remarks: First, note that subspaces with lower dimension can usually be more entangled than subspaces of higher dimension. However, here the dimension of the symmetric subspace is smaller than the dimension of the subspace with mixed symmetry, hence the hierarchy can not simply be explained by the dimensions of the subspaces. Second, although this hierarchy seems to hold in the special case of defining subspaces with respect to the symmetric group (note however, that this still needs to be shown for general number of particles $N$) there is no general statement that the maximally entangled state with respect to the GM always lies in the best replacement of the antisymmetric subspace. In fact, for example for five qubits the GM is maximized by the so-called ring cluster state \cite{SteinbergGuehne2024}, which is an absolutely maximally entangled graph state and cannot be related to these symmetry classes. It therefore remains an open question for further research to elaborate how maximally entangled states with respect to the GM can be identified in a fully analytical way.

Let us now come back to the three-qubit case. Following the example above, we know that for three particles there are three subspaces labeled by the partitions $[3,0,0],\,[2,1,0]$ and $[1,1,1]$. However, since the length of the partition cannot exceed the local dimension of the subspace, in the qubit case the last partition $[1,1,1]$ is not valid and therefore the partition $[2,1,0] = [2,1]$ is the best replacement \footnote{The totally antisymmetric subspace corresponds to the partition of the form $[1,\cdots,1]$ and the symmetric subspace to the partition $[N]$. We call that subspace \textit{the best replacement of the antisymmetric subspace}, which corresponds to the partition that has the maximum number nonzero entries, with each entry as small as possible.} of the antisymmetric one. This corresponds to the split $\mathcal{H} = \mathcal{H}_{\mathcal{S}}\oplus\mathcal{H}_{\mathcal{J}\oplus\bar{\mathcal{J}}}$. 

Now, instead of considering the full symmetric group, we now only consider a subgroup, namely the group of even permutations, called the \textit{alternating group}. This group has more irreps, which means that the subspace corresponding to the partition $[2,1]$ splits into two further subspaces. Interestingly, these are exactly the chiral and antichiral three qubit subspaces: $\mathcal{H} = \mathcal{H}_{\mathcal{S}}\oplus\mathcal{H}_{\mathcal{J}}\oplus\mathcal{H}_{\bar{\mathcal{J}}}$, as also discussed in \cite{WenWilczekZee1989}.
In fact, this additional splitting always happens if the corresponding partition is \textit{self conjugate}\footnote{To check whether a partition is self conjugate one can illustrate it in a \textit{Young diagram}. A Young diagram is a collection of left-oriented rows of boxes, where each entry $\lambda_i$ of a partition $\lambda$ denotes the number of boxes in that row. The partition $[2,1]$, for example, would be illustrated by one row of two boxes and one row of one box below. When flipping this Young diagram on its main diagonal it remains unchanged, hence the partition is self conjugate.}. 

Therefore, in order to generalize the chiral subspaces in that sense, we first consider the splitting into subspaces according to the Schur-Weyl duality with respect to the symmetric group. As mentioned before, each subspace is labeled by a partition $\lambda$ of $N$ particles. We then take the subspace, corresponding to a self conjugate partition. This subspace is the direct sum of our generalized $N$-particle chiral and antichiral subspace and to identify these we consider the splitting of that subspace according to the Schur-Weyl duality with respect to the alternating group. Note that the minimal local dimension for which these subspaces exist, is bounded by the length of the partition and therefore given by 
$d\geq \left\lceil\sqrt{N}\right\rceil$. The information needed to construct the projectors onto these subspaces can be found in the \textit{character table} of the alternating group of $N$ particles. Note, however, that constructing these projectors can become tedious for larger $N$. For a more illustrative and detailed explanation we refer to Ref.~\cite{ArnauAlbert}, where these subspaces were investigated in connection to parity detection.

Let us now consider some examples: For four qubits we find the self conjugate partition $\lambda = [2,2]$ and considering the irreps of the alternating group the corresponding subspace splits into two one-dimensional subspaces. These contain the Higuchi-Sudbery state \cite{HiguchiSudbery2000,BrierleyHiguchi2007} mentioned in the main text and its complex conjugate, which is known to be the maximally entangled state for four qubits \cite{DerksenFriedlandLimWang2017}. 

Since these are only one-dimensional subspaces, we further consider the 
five-particle case. 
The partition of interest is now $\lambda =[3,1,1]$, and therefore the local dimension must be at least three. We obtain two 18-dimensional five qutrit subspaces that are highly entangled. In fact, recently it was shown that the GM of states in these subspaces is at least $\nicefrac{17}{20}$ \cite{ArnauAlbert}, which is close to the highest possible value for five qutrits \cite{SteinbergGuehne2024} (see also Table~\ref{tab:res}). 

\begin{table}[t!!]
    \centering
    \begin{tabular}{|l|c|c|c|}\hline
         & LU construction & chirality operator & group theory \\\toprule
    number of particles     & $N = 4$ & $N = 5$ & $N = 5$ \\
    local dimension     & $d = 3$ & $d = 2$ & $d = 3$ \\
    subspace dimension     & $D = 3$ & $D = 6 $ & $D = 18$ \\\hline
     lower bound on the geometric measure & $\nicefrac{7}{8} = 0.8750$ & $0.5883$  & $\nicefrac{17}{20} = 0.8500$ \cite{ArnauAlbert}\\
     largest possible geometric measure \cite{SteinbergGuehne2024} & $\nicefrac{8}{9}\approx 0.8889$   &  $0.8686$  & $0.9612$ \\\hline
     condition for generalization & $d = N-1$ & $N$ odd, $d\geq 2 $ & $\left\lceil\sqrt{N}\right\rceil \leq d$ \\\hline
    \end{tabular}
    \caption{Summary of the three generalization methods. Number of particles $N$, local dimension $d$ and subspace dimension $D$ are given for the simplest nontrivial generalization. The bounds on the GM in the respective subspaces are computed via an SDP. 
    Lastly, we give the conditions for $N$ and $d$ under which the generalization is possible.}
    \label{tab:res}
\end{table}

\subsection{Entanglement witnesses for $N$ particles}
As seen before, there are several ways to generalize our results from the main text and identify highly-entangled subspaces. Surely, not only the spaces can be generalized, but also the entanglement witnesses. We recapitulate, that in the three particle case, the witness $W_-$ is based on the chirality operator $\chi$. As discussed before, the chirality operator can be defined for any odd $N$ and is given by \cite{chiralOpN}
    \begin{align}
        \chi_N 
        = \mathbf{i}[P_1,P_2],
    \end{align}
where we rescaled with a constant factor for better readability. The operators $P_i$ are permutation operators that can be constructed following Ref.~\cite{chiralOpN}. Precisely, for $i=1,\cdots, N$, $P_i$ maps qudit $i+j$ to position $(N+i-j) \text{ mod } N$. Expressed in terms of flip operators we can write
    \begin{align}
        P_1 = \Pi_{j=0}^{\lfloor\frac{N}{2}\rfloor}F_{(j+2)(N-j)},\label{eq:Pi}
    \end{align}
    and $P_2$ can be constructed from $P_1$ via $P_2=T_N^2P_1$. For example for $N=3$ we would have $P_1P_2 = T_3 = (T_3^\dagger)^2$ and $P_2P_1 = T_3^2$ and for $N=5$, we obtain $P_1P_2 = T_5^3 = (T_5^\dagger)^2$ and $P_2P_1 = T_5^2$.
We observe the following: \\

\begin{figure}[htbp]
    \centering
    \includegraphics[width=0.49\linewidth]{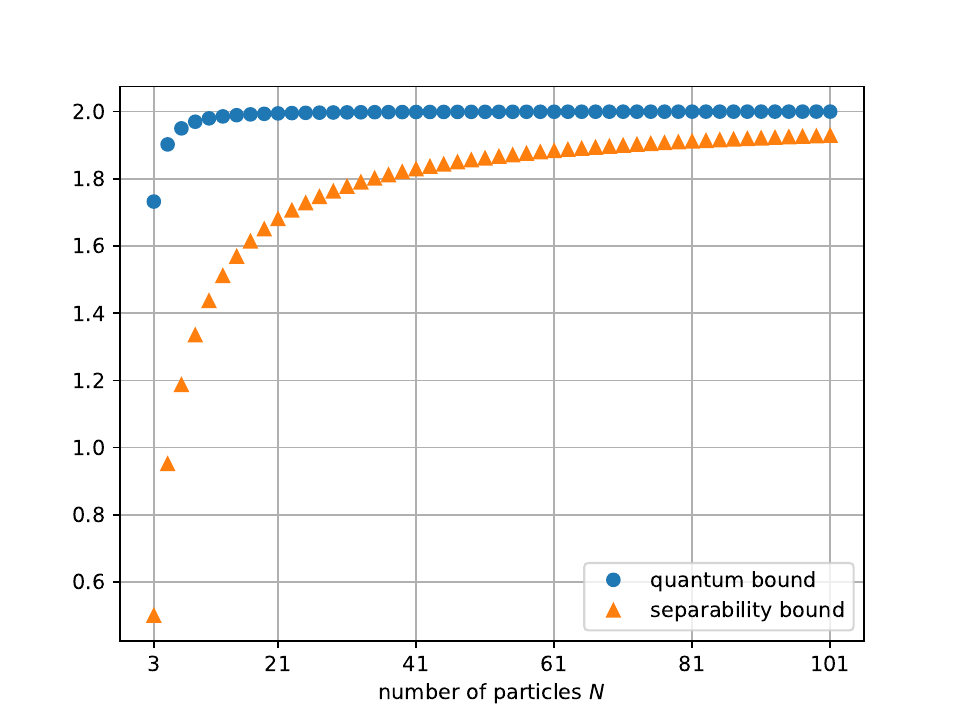}
    \includegraphics[width=0.49\linewidth]{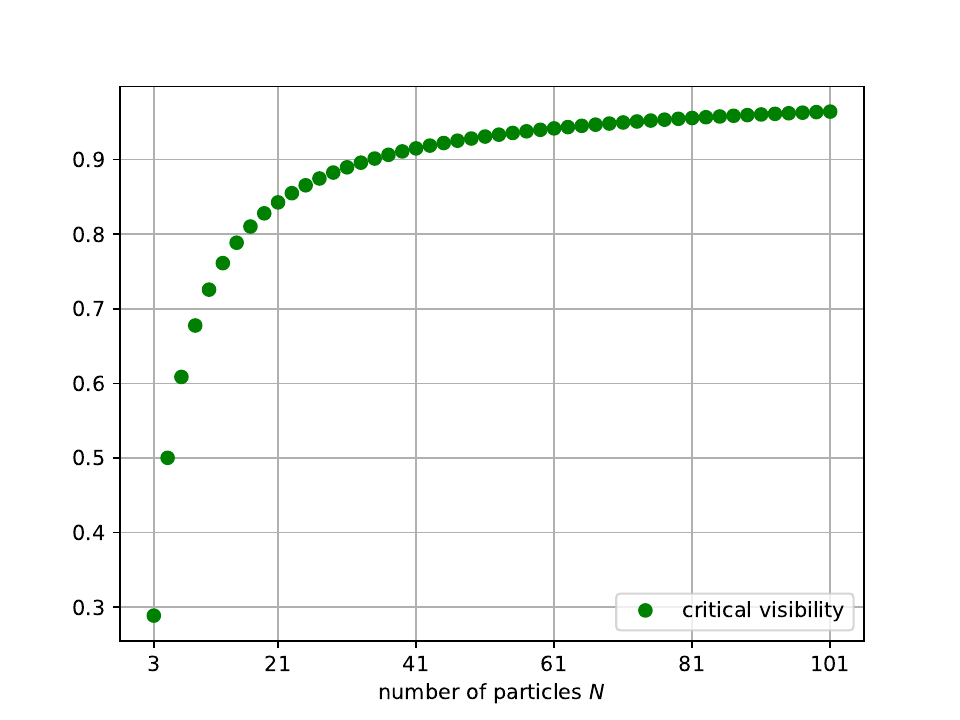}
    \caption{Left: Separability bounds compared to quantum bounds for the $N$ particle chirality operator $\chi_N$. Right: The critical visibility up to which the eigenvector corresponding to the largest eigenvalue of $\chi_N$ is still detected, when mixed with white noise. Note that in the plot the witness was renormalized, such that $\chi_N = \mathbf{i}[P_1,P_2]$.}
    \label{fig:Nwitness}
\end{figure}

\noindent
\textbf{Observation A1.} \textit{
The chirality operator, that generalizes the three particle operator $\chi_3 = \sum_{ijk}\varepsilon_{ijk}\sigma_i\otimes\sigma_j\otimes\sigma_k$ to $N$ particles, is a witness for multipartite entanglement. Precisely after rescaling $\chi_N$, it holds $\tr(\chi_N\sigma)\leq -2\mathrm{Im}(\eta_N)$ (as shown in Fig.~\ref{fig:Nwitness}), where $\eta_N = \bra{\pi}T_N^2\ket{\pi}$ is the expectation value of the squared $N$ particle cyclic translation operator $T_N$ over all product states $\ket{\pi}=\ket{a_1\cdots a_N}$ and can be interpreted as a Bargmann invariant of order $N$.
}
\begin{proof}
    Let us first consider the the commutator $[P_1,P_2]$ defining the chirality operator. Note here, that the $P_i$, which are defined in Eq.~\eqref{eq:Pi}, depend on $N$, but for better readability we omit this index. We have 
    \begin{align}
        [P_1,P_2] = P_1T_N^2P_1-T_N^2P_1P_1 = P_1T_N^2P_1 -T_N^2,
    \end{align}
    where we used that $P_1P_1=\mathds{1}$, since $P_1$ is a product of flip operators which act on distinct subsystems.

 For the term $P_1T_N^2P_1$ we have to consider the mapping again. For $i=1$ we have
    \begin{align}
        1+j\mapsto N+1-j \Leftrightarrow j \mapsto N+2-j 
    \end{align}
    and the operator $T_N^2$ maps it two positions further, so
    \begin{align}
        j\mapsto N+4-j = j'.
    \end{align}
    Now $P_1$ acts again on $j'$, yielding the total mapping
    \begin{align}
        j\mapsto N+2-(N+4-j) = j-2.
    \end{align}
    Therefore, we have $P_1T_N^2P_1 = T_N^{N-2} = (T_N^\dagger)^2$ and the chirality operator is given by
    \begin{align}
        \chi_N = \mathrm{i}[(T_N^\dagger)^2-T_N^2].
    \end{align}
    Now we can compute the overlap with product states:
    \begin{align}
        \tr(\chi_N\sigma)\leq \bra{\pi}\mathrm{i}((T_N^\dagger)^2-T_N^2)\ket{\pi} = \mathrm{i}(\eta_N^*-\eta_N)=\mathbf{i}[2\mathbf{i}\mathrm{Im}(\eta_N)]=-2\mathrm{Im}(\eta_N).
    \end{align}
    Here, we defined $\bra{\pi}T_N^2\ket{\pi}=:\eta_N$, where $\ket{\pi}=\ket{a_1\cdots a_N}$ is an $N$-partite product state. Such expressions $\eta_N$ are so-called \textit{Bargmann invariants} \cite{Bargmann,Bargmann2,Bargmannrev}. In general, a Bargmann invariant of order $N$ is defined as the multivariate trace of $N$ states $\varrho_i$, $\tr(\varrho_1\cdots\varrho_N)$. Taking the $\varrho_i = \ket{a_i}\bra{a_i}$ as pure states, this yields, for example if $N=3$: $\tr(\ket{a_1}\braket{a_1}{a_2}\braket{a_2}{a_3}\bra{a_3}) = \braket{a_1}{a_2}\braket{a_2}{a_3}\braket{a_3}{a_1} = \bra{a_1}\bra{a_2}\bra{a_3}T_3^2\ket{a_1}\ket{a_2}\ket{a_3}=\eta_3$.
    Recently, in Refs.~\cite{ChaBarkmann,ChaBarkmann2} the values Bargmann invariants of order $N$ may attain, were investigated. They are given by
    \begin{align}
        \eta_N = -\left(\frac{\cos(\pi/N)}{\cos(\theta/N)}\right)^Ne^{\mathbf{i}\theta}
    \end{align}
    for $\theta\in [-\pi,\pi[$. Since we are interested in maximizing the negative imaginary part of $\eta_N$ we have to maximize:
    \begin{align}
        -2\mathrm{Im}(\eta_N) \leq \max_\theta \left[ 2\left(\frac{\cos(\pi/N)}{\cos(\theta/N)}\right)^N\sin(\theta)\right]\label{eq:sepbound}
    \end{align}
     and after some algebra, we find that the expression is maximized for $\theta = \frac{\pi}{2}\frac{N}{N-1}$, allowing us to analytically determine the separability bound of the witness for all $N$. Note, that due to the rescaling of the operator $\chi_N$ and the fact that Bargmann invariants are defined for any dimension, the witness is dimensionless, meaning that this bound holds for all local dimensions $d$.
\end{proof}
Let us now discuss the robustness of these witnesses. To that end, we first determine the quantum bound, which is given by the operator's largest eigenvalue. Since $\chi_N$ is a polynomial in the translation operator $T_N$, $\chi_N$ and $T_N$ have common eigenvectors and we can directly compute the eigenvalues $\lambda_j$ of $\chi_N$. We obtain
\begin{align}
   \lambda_j = \mathbf{i}\omega_N^{(N-2)j}-\mathbf{i}\omega_N^{2j}= \mathbf{i}\omega_N^{-2j}-\mathbf{i}\omega_N^{2j} = 2\sin(\frac{4\pi j}{N}),\label{eq:maxeig}
\end{align}
where $\omega_N^j = e^{\frac{2\pi\mathbf{i}}{N}j}$, for $j=1,\cdots N$ are the eigenvalues of $T_N$. 
We can therefore directly compute the quantum bound which is plotted and compared to the separability bound in Fig.~\ref{fig:Nwitness}.

In order to determine the robustness of our witnesses, we take a look at their sprectra to find potential candidates for detectable states. Again, since $\chi_N = \mathbf{i}[(T_N^\dagger)^2-T_N^2]$ is a polynomial in $T_N$, the operators commute and hence have common eigenvectors. These correspond to the Fourier basis and in the qubit case are given by
\begin{align}
    \ket{\psi_j^N} = \frac{1}{\sqrt{N}}\sum_{k=1}^N\omega_N^{-jk}T_N^k\ket{0\cdots 01},
\end{align}
where the index $j$ denotes the corresponding eigenvalue $\lambda_j$, depending on $N$ as defined in Eq.~\eqref{eq:maxeig}. These states are so-called \textit{phased Dicke states} \cite{phase1,phase2,phase3,phase4,phase5,phase6,phase7}; we therefore recover and generalize the finding in the three-qubit case \cite{ImaiTothGuehne2024}.
Note that a generalization to qudits is possible in analogy to the 
three-particle case discussed in the main text.

We can now compute the critical visibility of the witnesses, i.e. the minimal mixing parameter such that the target state is still detected when mixed with noise. In this case we choose one of the phased Dicke states, corresponding to the largest eigenvalue as target state and mix it with white noise
\begin{align}
    \rho = p\ket{\psi_{j^*}^N}\bra{\psi_{j^*}^N}+(1-p)\frac{\mathds{1}}{d^N}.
\end{align}
Since $\chi_N$ is traceless, the critical visibility $p_\mathrm{crit}$ is then given by
\begin{align}
    p_\mathrm{crit} = \frac{\alpha}{\lambda_{j^*}},
\end{align}
where $\alpha$ is the separability bound, given in Eq.~\eqref{eq:sepbound} and $\lambda_{j^*}$ the largest eigenvalue of $\chi_N$, which can be computed using Eq.~\eqref{eq:maxeig}. The results are shown in Fig.~\ref{fig:Nwitness}. We see that the witness does not perform well, when the number of particles increases. In fact, for very large $N$ the separability bound will coincide with the quantum bound making it impossible to detect any entanglement. We note however, that here we just give a proof of principle using a rather simple construction for the witness operator. Finding more complex but better performing witnesses is left open for future studies.

\section{Proof of Observations 1 and 2}
In this appendix we show that the geometric measure of entanglement is lower bounded for the (anti)chiral symmetric subspaces and their direct sum. We start by proving the bound for the direct sum of the two subspaces, then we consider the chiral subspace, which can be treated analogously to the antichiral subspace. Lastly, we prove two Lemmas that are necessary to calculate these bounds and show that they hold for any local dimension $d$.

\subsection{Bound for the direct sum of the chiral and antichiral subspace}\label{sub:JplusJ}
\subsubsection{The qubit case (Observation 1)}
We will now give a proof of the first part of Observation~1.

\noindent 
\textbf{Observation 1a.}\textit{
All states in the direct sum of the chiral and antichiral space
$\mathcal{H}_{\mathcal{J}}\oplus \mathcal{H}_{\Bar{\mathcal{J}}}$
have an entanglement of at least $G(\ket{\psi})\geq\nicefrac{1}{4}.$
}
\begin{proof}
First, recall that the projectors onto all subspaces sum up to one: $\Pi_\mathcal{S} + \Pi_\mathcal{J} + \Pi_{\mathcal{\Bar{J}}}=\mathds{1}$. Then, we apply the following logic to obtain an estimation of the lower bound of the geometric measure of entanglement (GM) 
\begin{align}
    G(\ket{\psi}) &= 1- \max_{\ket{abc}}|\braket{abc}{\psi}|^2\geq 1 - \max_{\ket{abc}} \bra{abc}\Pi_\mathcal{J} + \Pi_\mathcal{{\Bar{J}}}\ket{abc}= \min_{\ket{abc}} \bra{abc}\Pi_\mathcal{S}\ket{abc}.
\end{align}
Hence, we need to minimize the expectation value of the projector onto the symmetric subspace for all product vectors:
\begin{align}
    \bra{abc}\Pi_\mathcal{S}\ket{abc} &= \bra{abc}\frac{1}{3}\left(\mathds{1}+T+T^2\right)\ket{abc} = \frac{1}{3}\left(1+\bra{abc}T\ket{abc}+\bra{abc}T^2\ket{abc}\right)= \frac{1}{3}\left(1+ \eta + \eta^*\right)=\frac{1}{3}\left(1+2\mathrm{Re}(\eta)\right),\label{eq:pisymm}
\end{align}
where we defined $\eta \coloneqq \bra{abc}T\ket{abc}$. Here again, $\eta$ is a Bargmann invariant \cite{Bargmann,Bargmann2,Bargmannrev}, so we could determine the maximum real part directly, using Refs.~\cite{FernandesWagneretal,ChaBarkmann,ChaBarkmann2}. However, in the end of this appendix, we will give an alternative proof specifically for the real part of $\eta$.\\

\noindent
\textbf{Lemma 1.} \textit{The expectation value of the cyclic shift operator $T$ for product states $\ket{abc}$ fulfills $\mathrm{Re}(\eta) {= \mathrm{Re}(\bra{abc}T\ket{abc}) \geq -\nicefrac{1}{8}}$.}

With this Lemma we have
\begin{align}
    \bra{abc}\Pi_\mathcal{S}\ket{abc} &= \frac{1}{3}\left(1+2\mathrm{Re}(\eta)\right) \geq \frac{1}{3}\left(1-2\times\frac{1}{8}\right)= \frac{1}{4}.
\end{align}
Hence, for all pure states in the direct sum of the chiral and antichiral subspace $\mathcal{H_J}\oplus\mathcal{H}_{\Bar{\mathcal{J}}}$ it holds $G(\ket{\psi})\geq \nicefrac{1}{4}$.
\end{proof}

Note that in the qubit case the result $\bra{abc}\Pi_\mathcal{S}\ket{abc}\geq \nicefrac{1}{4}$ recovers the findings in \cite{DAlessandro2023}, where the overlap of product states with projectors onto the symmetric subspace was calculated.

\subsubsection{The higher-dimensional case $d\geq 3$ (Observation 2)}
We now want to consider the bound of the geometric measure of entanglement for any local dimension $d\geq 3$.

\noindent
\textbf{Observation 2a.}\textit{
In higher-dimensional systems ($d\geq 3$), all states in the direct sum of the chiral and antichiral space
$\mathcal{H}_{\mathcal{J}}\oplus \mathcal{H}_{\Bar{\mathcal{J}}}$
have an entanglement of at least $G(\ket{\psi})\geq\nicefrac{1}{4}.$
}
\begin{proof}
In this case there exist not only the symmetric, chiral and antichiral subspace, but also an antisymmetric subspace. So, there are projectors $\Pi_\mathcal{S}+ \Pi_\mathcal{A} + \Pi_\mathcal{J} + \Pi_{\Bar{\mathcal{J}}} = \mathds{1}$.
However, we see that the projector onto the direct sum of the symmetric and antisymmetric subspace has the same form as the projector onto the symmetric subspace in the qubit case (see also Eq.~\ref{eq:pisymm})
\begin{align}
    \Pi_{\mathcal{S}\oplus \mathcal{A}} = \frac{1}{6}\left(\mathds{1} + F_{12} + F_{23} + F_{13} + T + T^2\right) +\frac{1}{6}\left(\mathds{1} - F_{12} - F_{23} - F_{13} + T + T^2\right)
    = \frac{1}{3}(\mathds{1}+T+T^2).
\end{align}
Thus, to compute the bound for the geometric measure of entanglement we can proceed as above. Since Lemma~1 is valid for all dimensions $d$, as we will see in the last section of this appendix, we conclude that for all states in the direct sum of the chiral and antichiral symmetric subspace in any local dimension $d$ the GM fulfills $G(\ket{\psi})\geq\nicefrac{1}{4}$.
\end{proof}

\subsection{Bound for the (anti)chiral subspace}\label{sub:J}
We now show that the GM for all states in the (anti)chiral subspace is lower bounded by $\nicefrac{5}{9}$.

\noindent
\textbf{Observation 1b \& Observation 2b.}\textit{
For qubit ($d=2$) and higher-dimensional systems 
($d \geq 3$),  any pure state in the chiral space 
$\mathcal{H}_{\mathcal{J}}$  (or, equivalently, the antichiral 
space $\mathcal{H_{\Bar{\mathcal{J}}}}$) has a geometric measure
of at least $G(\ket{\psi}) \geq \nicefrac{5}{9}$.
}
\begin{proof}
Let us consider the chiral symmetric subspace. The computation for the antichiral symmetric subspace is analogous.
Following the same logic as above, we now have to minimize the expectation value of the complementary projectors
\begin{align}
    \bra{abc}\Pi_\mathcal{S} +\Pi_\mathcal{A} + \Pi_{\Bar{\mathcal{J}}}\ket{abc} &= \bra{abc}\frac{1}{3}\left(\mathds{1}+T+T^2\right)+\frac{1}{3}\left(\mathds{1}+\omega T + \omega^2 T^2\right)\ket{abc}\notag\\
    &= \frac{1}{3}\left[2+(\omega+1)\eta+(\omega^*+1)\eta^*\right]= \frac{1}{3}\left[ 2+2 {\rm Re}(-\omega^2\eta)\right],
\end{align}
where we used the fact that $\Pi_\mathcal{S} +\Pi_\mathcal{A} = \nicefrac{1}{3}\left(\mathds{1}+T+T^2\right)$, $\omega+\omega^2+1=0$, and $\omega^2=\omega^*$. Now, we apply another Lemma, which we will prove in the next subsection:

\noindent
\textbf{Lemma 2:} \textit{The real part of the expectation value of the cyclic shift operator $T$ for product states $\ket{abc}$ multiplied with the phase $e^{\mathbf{i}\frac{\pi}{3}}$ fulfills ${\rm Re}(-\omega^2\eta) = {\rm Re}(e^{\mathbf{i}\frac{\pi}{3}}\eta)\geq-\frac{1}{6}$.}

Then, we obtain
\begin{align}
    \bra{abc}\Pi_S + \Pi_{\Bar{J}}\ket{abc} &\geq \frac{1}{3}\left(2-2\times \frac{1}{6}\right) =\frac{5}{9}.
\end{align}
and knowing that $\nicefrac{5}{9}$ is the largest possible GM in the three qubit case \cite{TamaryanWeiPark2009}, we showed that for any state in the chiral symmetric subspace $\mathcal{H_J}$ the geometric measure of entanglement fulfills $G(\ket{\psi})=\nicefrac{5}{9}$.  

Note that with the same argument as above and since Lemma~2 is independent of the dimension as well, we see that the value $\nicefrac{5}{9}$ is a lower bound for any local dimension $d$.
\end{proof}

\subsection{Proof of Lemma~1 and Lemma~2}
We now want to give the proof of the previously introduced Lemma~1 and Lemma~2. More precise, we want to show the following:
\begin{align}
    \mathrm{Re}(e^{\mathbf{i}\alpha}\eta) \geq \left\{ \begin{matrix}
    -\nicefrac{1}{8}\,\,\text{for}\,\,\alpha=0\\
    -\nicefrac{1}{6}\,\,\text{for}\,\,\alpha=\frac{\pi}{3}
    \end{matrix}\right\},
\end{align}
for $\eta = \bra{abc}T\ket{abc}$.
\begin{proof}
We minimize
\begin{align}
    \mathrm{Re}(\bra{abc}T\ket{abc}e^{\mathbf{i}\alpha}) &= \mathrm{Re}(\braket{abc}{cab}e^{\mathbf{i}\alpha}) 
    = \mathrm{Re}(\braket{a}{c}\braket{b}{a}\braket{c}{b}e^{\mathbf{i}\alpha})
    = \mathrm{Re}(\braket{a}{c}\braket{c}{b}\braket{b}{a}e^{\mathbf{i}\alpha})\notag\\
    &= \mathrm{Re}(\bra{a}\left(\frac{1}{2} (| c\rangle\langle c\ket{b}\bra{b} e^{\mathbf{i}\alpha} + (| c\rangle\langle c\ket{b}\bra{b})^\dagger e^{-\mathbf{i}\alpha})\right)\ket{a})
    = \bra{a}M\ket{a},\label{eq:proof2}
\end{align}
which corresponds to finding the smallest eigenvalue of the hermitian matrix $M = \nicefrac{1}{2} ((| c\rangle\langle c\ket{b}\bra{b} e^{\mathbf{i}\alpha} + (| c\rangle\langle c\ket{b}\bra{b})^\dagger e^{-\mathbf{i}\alpha})$.

Let us start considering qubits, since later we will see that the calculation in any local dimension will reduce to this case.
As $T$ is $U^{\otimes 3}$-invariant, without loss of generality, we can write $\ket{c}=\ket{0}$ and $\ket{b} = \cos(\theta)\ket{0}+e^{\mathbf{i}\varphi}\sin(\theta)\ket{1}$. This yields
\begin{align}
    M = \frac{1}{2}\cos(\theta) 
    \left(\begin{matrix}
    \cos(\theta)(e^{\mathbf{i}\alpha}+e^{-\mathbf{i}\alpha}) & \sin(\theta)e^{\mathbf{i}(\alpha-\varphi)}\\
    \sin(\theta)e^{-\mathbf{i}(\alpha-\varphi)} & 0
    \end{matrix}\right).
\end{align}
And the eigenvalues are given by
\begin{align}
    \lambda_\pm = \frac{1}{2}\cos(\theta)\left(\frac{C\cos(\theta)}{2}\pm\sqrt{\left(\frac{C\cos(\theta)}{2}\right)^2+\sin(\theta)^2}\right),\label{eq:mineval}
\end{align}
where $C \coloneqq 2\cos(\alpha)$. 

Let us consider the first case, where $\alpha = 0$ and therefore $C = 2$. Then Eq.~(\ref{eq:mineval}) reduces to
\begin{align}
    \lambda_\pm &= \frac{1}{2}\cos(\theta)\left( \cos(\theta)\pm\sqrt{\cos(\theta)^2+\sin(\theta)^2}\right)= \frac{1}{2} \cos(\theta)(\cos(\theta) \pm 1)
\end{align}
and becomes minimal for $\cos(\theta) = \mp\nicefrac{1}{2}$, reaching the value $\lambda_\mathrm{min} = -\nicefrac{1}{8}$, which indeed coincides with the bound given in Refs.~\cite{ChaBarkmann,ChaBarkmann2}:
\begin{align}
    \mathrm{Re}(\eta) \geq \lambda_\mathrm{min} = -\frac{1}{8}.
\end{align}
In the second case, where $\alpha=\nicefrac{\pi}{3}$ and $C=1$, we obtain a minimum for $\cos(\theta)=\mp\nicefrac{1}{\sqrt{3}}$ and it follows:
\begin{align}
    \mathrm{Re}(e^{\mathbf{i}\frac{\pi}{3}}\eta) \geq \lambda_\mathrm{min} = -\frac{1}{6}.
\end{align}
Hence, the three-qubit case is done. It remains to show that this calculation is independent of the local dimension. 

Consider again the expression $\mathrm{Re}(\eta e^{\mathbf{i}\alpha})=\mathrm{Re}(\bra{abc}T\ket{abc}e^{\mathbf{i}\alpha})$ we want to bound. Then, following the calculations in Eq.~(\ref{eq:proof2}) we obtain
\begin{align}
    \mathrm{Re}(\bra{abc}T\ket{abc}e^{\mathbf{i}\alpha})= \mathrm{Re}(\bra{a}\left(\frac{1}{2} (|c\rangle \braket{c}{b}\langle b| e^{\mathbf{i}\alpha} + (|c\rangle \braket{c}{b}\langle b|)^\dagger e^{-\mathbf{i}\alpha})\right)\ket{a}) = \bra{a}M_{d\times d}\ket{a}. 
\end{align}
Since we now consider particles of local dimension $d\geq 3$, we have a different parameterization for $\ket{b}$ and $\ket{c}$. However, we can still write $\ket{c} = \ket{0}$ and $\ket{b} = \cos(\theta)\ket{0}+e^{\mathbf{i}\varphi}\sin(\theta)\ket{\psi_c}$, with some normalized state $\ket{\psi_c}$, which is orthogonal to $\ket{0}$. Then the matrix $M_{d\times d}$ reads as follows
\begin{align}
    M_{d\times d} = \frac{1}{2}\cos(\theta) 
    \left(\begin{matrix}
        \cos(\theta)(e^{\mathbf{i}\alpha}+e^{-\mathbf{i}\alpha}) & e^{\mathbf{i}(\alpha-\varphi)}\sin(\theta) \bra{\psi_c}\\
        e^{-\mathbf{i}(\alpha-\varphi)}\sin(\theta)\ket{\psi_c} & \mathbf{0}_{(d-1)\times(d-1)}
    \end{matrix}\right).
\end{align}
We note that $M_{d\times d}$ is a block matrix of the following form
\begin{align}
    M_{d\times d} = \left(\begin{matrix}
        c_0 & c_1 & \cdots & c_{d-1}\\
        c_1^* & 0 & \cdots & 0 \\
        \vdots & \vdots & \ddots & \vdots \\
        c_{d-1}^* & 0 & \cdots & 0 
    \end{matrix}\right).
\end{align}
Therefore, we can apply Schur's formula to compute the determinant yielding the characteristic polynomial. Schur's formula for determinants of a block matrix $L$ is given by \cite{Schur1917}:
\begin{align}
    \det(L)=\det(D)\det \left(A-BD^{-1}C\right) = \det(A)\det \left(D-CA^{-1}B\right). 
\end{align}
For $L= M_{d\times d}-\lambda\mathds{1}$ we have $A = c_0-\lambda$, $B = (c_1,...,c_{n-1})$, $C = (c_1^*,...,c_{n-1}^*)^T$ and $D = -\lambda\mathds{1}_{(d-1)\times (d-1)}$. This yields
 \begin{align}
    \det(M_{d\times d}-\lambda\mathds{1}) &= \det(D)\det \left(A-BD^{-1}C\right)= (-\lambda)^{d-1}\det(c_0-\lambda + \frac{1}{\lambda}\sum_i|c_i|^2)
    = (-\lambda)^{d-1}(c_0-\lambda) - (-\lambda)^{d-2}\sum_i|c_i|^2.
 \end{align}
We see that $d-2$ eigenvalues are equal to zero. The only two nonzero eigenvalues are given by
\begin{align}
    \lambda_\pm &= \frac{c_0}{2}\pm\sqrt{\left(\frac{c_0}{2}\right)^2+\sum_i|c_i|^2}
    = \frac{1}{2}\cos(\theta)\left(\frac{2\cos(\alpha)\cos(\theta)}{2}\pm\sqrt{\left(\frac{2\cos(\alpha)\cos(\theta)}{2}\right)^2+\sin(\theta)^2|\braket{\psi_c}{\psi_c}|^2}\right),
\end{align}
with $C \coloneqq 2\cos(\alpha)$ this reduces to Eq.~(\ref{eq:mineval}) in the qubit case, since $\ket{\psi_c}$ is normalized, so $|\braket{\psi_c}{\psi_c}|^2 =1$. Thus, for any local dimension $d$ it holds $\lambda_\text{min} = -\nicefrac{1}{8}$ if $\alpha = 0$ (Lemma~1) and $\lambda_\text{min} = -\nicefrac{1}{6}$ if $\alpha = \nicefrac{\pi}{3}$ (Lemma~2). 
\end{proof}

\section{Generalization of Bell state measurements for three qubits}
In this appendix we will show how the projectors onto the symmetric and (anti)chiral symmetric subspaces can be used as effects of a POVM to estimate
quantities such as $\tr(\varrho^3)$.

First, recapitulate that a positive operator valued measurement (POVM) is a set of positive operators $\{O_i\}$, that sum up to one: $\sum_iO_i=\mathds{1}$. These operators are called the \textit{effects} of the POVM. Indeed the projectors $\Pi_\mathcal{S}, \Pi_\mathcal{J}$ and $\Pi_{\mathcal{\Bar{J}}}$ are positive operators and fulfill $\Pi_\mathcal{S} + \Pi_\mathcal{J} + \Pi_{\mathcal{\Bar{J}}} = \mathds{1}$. They can therefore be interpreted as effects of a POVM. We now show how this POVM can be used to estimate $\tr(\varrho^{3})$.

Let us consider the corresponding circuit \cite{Bargmanncircuit1, Bargmanncircuit2,LiuKnoerzerWangTura2024}:
\begin{align}
    \begin{quantikz}
\lstick{$\ket{z}_a$} & \gate{\mathcal{F}} & \ctrl{1}  & \gate{\mathcal{F}^\dagger} &\qw \\
\lstick{$\ket{\phi}$} & \qw & \gate{\mathcal{D}} &  \qw & \qw
\end{quantikz}\label{circuit}
\end{align}    
where the inputs are an ancilla $\ket{z}_a$ in three dimensions and a three-qubit state $\ket{\psi}$. The gate $\mathcal{F}$ is the Fourier transform on a qutrit and given by
\begin{align}
    \mathcal{F} = \frac{1}{\sqrt{3}}\left(\begin{matrix}
        1 & 1 & 1\\
        1 & \omega & \omega^2\\
        1 & \omega^2 &\omega
    \end{matrix}\right),
\end{align}
with $\omega = e^{\frac{2\pi \mathbf{i}}{3}}$
and $\mathcal{D}$ is a controlled gate (we write $C\mathcal{D}$) that applies
\begin{align}
    &\mathds{1}\,\,\text{if the ancilla is in}\,\, \ket{0}_a,\\
    &T^2\,\,\text{if the ancilla is in}\,\, \ket{1}_a,\\
    &T\,\,\text{if the ancilla is in}\,\, \ket{2}_a.
\end{align}
Then, if we measure $\ket{1}$ on the ancilla system, we obtain:
\begin{align}
    \bra{1}_a\mathcal{F}^\dagger (C\mathcal{D})\mathcal{F}\ket{0}_a\otimes \ket{\psi} =& \frac{1}{3}(\bra{0}_a+\omega^2\bra{1}_a+\omega\bra{2}_a)C\mathcal{D}(\ket{0}_a+\ket{1}_a+\ket{2}_a)\otimes\ket{\psi}\notag \\
    =& \frac{1}{3}(\bra{0}_a+\omega^2\bra{1}_a+\omega\bra{2}_a)(\ket{0}_a\otimes\mathds{1}\ket{\phi}+\ket{1}_a\otimes T^2\ket{\phi}+\ket{2}_a\otimes T\ket{\psi})\notag\\
    =& \frac{1}{3}(\mathds{1} + \omega^2T^2 + \omega T)\ket{\psi} = \Pi_\mathcal{\bar{J}}\ket{\psi}.
\end{align}
This corresponds to a projection on the antichiral symmetric subspace. Analogously, measuring $\ket{0}_a$ and $\ket{2}_a$ yield a projection onto the symmetric and chiral symmetric subspace, respectively.

It follows that the probability to obtain outcome $\ket{i}$ is given by
\begin{align}
    p(i) = \braket{\tilde{\psi}}{\tilde{\psi}} = \bra{\psi}\Pi_i\Pi_i\ket{\psi} = \bra{\psi}\Pi_i\ket{\psi},
\end{align}
where $\Pi_i \in \{\Pi_\mathcal{S},\Pi_\mathcal{\bar{J}},\Pi_\mathcal{J}\}$, for $i=0,1,2$, respectively. Hence, we showed that measuring the POVM $\{\Pi_\mathcal{S},\Pi_\mathcal{J},\Pi_\mathcal{\bar{J}}\}$ is equivalent to applying the circuit~(\ref{circuit}) and can therefore be used to estimate quantities like $\tr(\varrho^3)$, which correspond to Bargmann invariants and moreover are useful for entanglement measures such as the generalized concentratable entanglement~\cite{LiuKnoerzerWangTura2024}.

Note that this approach generalizes the SWAP-test which can be implemented by a maximally entangled measurement on two qubits \cite{BeckeyPelegríFouldsPearson2022}. See further Refs.~\cite{Bargmanncircuit1, Bargmanncircuit2} for similar results.

%%%%%%%%%%%%%%%%%%%%%%%%%%%%%%%%%%%%%%%%%%%%%%%%%%%%%%%%%%%%%
\section{Proofs of the properties of the observables $W_\pm$
in the main text}
\subsubsection{Invariance of the constructing operators}

\noindent
\textbf{Proposition 1.}\textit{
The operators $W_\pm = \sum_{ijk}\kappa_{ijk}^\pm \lambda_i\otimes\lambda_j\otimes\lambda_k$, with $\kappa^+_{ijk}=\nicefrac{1}{d^2}\tr(\lambda_i\{\lambda_j,\lambda_k\})$ and $\kappa^-_{ijk}=-\nicefrac{\mathbf{i}}{d^2}\tr(\lambda_i[\lambda_j,\lambda_k])$
are independent of the choice of the constructing basis operators $\lambda_i$. In fact, for any arbitrary set of orthonormal, traceless and hermitian operators $\{G_i\}$, with normalization $\tr(G_iG_j)=d\delta_{ij}$, it holds $\tilde{W}_-=\sum_{ijk}-\nicefrac{\mathbf{i}}{d^2}\tr(G_i^A[G^B_j,G^C_k])G^A_i\otimes G^B_j\otimes G^C_k = W_-$ and analogously for $W_+$. 
}
\begin{proof}
By orthogonal transformations $\lambda_j \mapsto G_i = \sum_j O_{ij}\lambda_j$, we can reach any ONB of hermitian and traceless operators $\{G_i\}$. Now, let us consider the expression $W = \sum_{ijk} \tr(\lambda_i\lambda_j\lambda_k) (\lambda_i\otimes\lambda_j\otimes\lambda_k)$. After performing an arbitrary basis change on parties $A,B$ and $C$ we have:
\begin{align}
    \tilde{W} &= \sum_{ijk} \left(\tr\left(\sum_{\alpha}O^A_{i\alpha }\lambda_\alpha\sum_\beta O^B_{j\beta}\lambda_\beta\sum_{\gamma}O^C_{k\gamma}\lambda_\gamma\right) \sum_{\alpha '\beta '\gamma '}(O^A_{i\alpha '}\lambda_{\alpha'}\otimes O^B_{j\beta '}\lambda_{\beta'}\otimes O^C_{k\gamma '}\lambda_{\gamma'})\right)\notag\\
    &= \sum_{\alpha\beta\gamma}\sum_{\alpha'\beta'\gamma'}\delta_{\alpha\alpha'}\delta_{\beta\beta'}\delta_{\gamma\gamma'}\tr(\lambda_\alpha\lambda_\beta\lambda_\gamma)(\lambda_{\alpha'}\otimes\lambda_{\beta'}\otimes\lambda_{\gamma'})
    =\sum_{\alpha\beta\gamma}\tr(\lambda_\alpha\lambda_\beta\lambda_\gamma)(\lambda_\alpha\otimes\lambda_\beta\otimes\lambda_\gamma) = W,
\end{align}
which yields the same expression $W$ previous to the basis change.
\end{proof}
\subsubsection{$U^{\otimes 3}$-invariance and Schur-Weyl duality}

\noindent
\textbf{Proposition 2.}\textit{
The operators $W_\pm$ are $U\otimes U\otimes U$-invariant and commute.
}\\
\begin{proof}
We first show the $U\otimes U\otimes U$-invariance. This follows from the invariance of the constructing operators.
Again, it is sufficient to consider the simplified operator $W$.
\begin{align}
    &U\otimes U\otimes U W U^\dagger \otimes U^\dagger\otimes U^\dagger = \sum_{ijk} \tr(\lambda_i\lambda_j\lambda_k) U\lambda_iU^\dagger\otimes U\lambda_jU^\dagger\otimes U\lambda_k U^\dagger \notag\\
    = &\sum_{ijk} \tr(U\lambda_iU^\dagger U\lambda_jU^\dagger U\lambda_kU^\dagger)U\lambda_iU^\dagger\otimes U\lambda_jU^\dagger\otimes U\lambda_k U^\dagger 
    = \sum_{ijk}\tr(G_iG_jG_k)G_i\otimes G_j\otimes G_k = W,
\end{align}
where in the first step we used the cyclicity of the trace and in the second the fact that any unitary transformation simply denotes a basis change: $U\lambda_iU^\dagger \mapsto G_i$. It follows that the operators $W_\pm$ are invariant under $U\otimes U\otimes U$-transformations, which completes the first part of the proof.

From the Schur-Weyl duality \cite{Weyl1946} it follows that any $U\otimes U\otimes U$-invariant operator can be expressed as a linear combination of permutation operators, such as the flip operators $F_{ij}$ or the cyclic and anticyclic translation operators $T$ and $T^2$. Hence, also for the operators $W_\pm$ exists such a decomposition. To find this we use the Weingarten calculus \cite{CollinsMatsumotoNovak2022,Koestenberger2021} and obtain
\begin{align}
    W_- &= \mathbf{i}\times d(T-T^2)\label{eq:decWA}\\
    W_+ &= d(T+T^2)-2(F_{12}+F_{23}+F_{13})+\frac{4}{d}\mathds{1}\label{eq:decWS}.
\end{align}
In this representation, we find that these operators are connected to the work in Ref.~\cite{EggelingWerner2001}, which investigates separability properties of $U\otimes U\otimes U$-invariant states.

This decomposition now allows us to show the commutation property.

Let us consider the following properties of the permutation operators:
\begin{align}
    F_{12}T = F_{23} = F_{13}T^2,\,\,
    F_{23}T = F_{13} = F_{12}T^2,\,\,
    F_{13}T = F_{12} = F_{23}T^2,\,\,
    TT^2 = T^2T = \mathds{1}.
\end{align}
Using these, we can compute that $W_+W_- = W_-W_+$ and thus $[W_+,W_-]=0$.
\end{proof}

\noindent
\textbf{Remark.} The fact that $W_\pm$ commute implies that they have common eigenvectors. We can write $W_+ = c_\mathcal{S}\Pi_\mathcal{S}+c_\mathcal{A}\Pi_\mathcal{A}+c_{\mathcal{J}}\Pi_{\mathcal{J}\oplus\bar{\mathcal{J}}}$ and $W_- = \alpha(\Pi_{\bar{\mathcal{J}}}-\Pi_\mathcal{J})$, where the coefficients $\alpha,c_\mathcal{A},c_\mathcal{S}$ and $c_\mathcal{J}$ follow from Eqs.~(\ref{eq:decWA}) and (\ref{eq:decWS}).

\section{Proof of Observation 4}

In the main text, we already saw that the operators' expectation values fulfill the following bounds for fully separable (fs), biseparable (bs) and general quantum states (q):
\begin{align}
    \langle W_-\rangle &\overset{\text{fs}}{\leq} \frac{d}{2}\overset{\text{bs}}{\leq}d\overset{\text{q}}{\leq}d\sqrt{3},\,\,d\geq 3\\
    \langle W_+\rangle &\overset{\text{fs}}{\leq} \frac{4}{3}\overset{\text{bs}}{\leq}\frac{10}{3}\overset{\text{q}}{\leq}\frac{40}{3},\,\,d=3\\
    \langle W_+\rangle &\overset{\text{fs\&bs}}{\leq} \frac{2(d-1)(d-2)}{d}\overset{\text{q}}{\leq}\frac{2(d+1)(d+2)}{d},\,\,d\geq 4.
\end{align}
We will now prove these bounds one by one. Let us start with the separability bound for $W_-$.

\noindent
\textbf{Observation 4a.}\textit{
    The witness operator based on the commutator fulfills the bound $\tr(W_-\sigma) \leq \nicefrac{d}{2}$ for all fully separable states~$\sigma$.}
\begin{proof}
From Eq.~(\ref{eq:decWA}) we know that $W_-\propto \mathbf{i}(T-T^2)$, so we consider:
\begin{align}
    \mathbf{i}\bra{abc}T-T^2\ket{abc} = \mathbf{i}(\braket{abc}{cab}-\braket{abc}{bca}) = \mathbf{i}(\eta-\eta^*) = \mathbf{i}[2\mathbf{i}\textrm{Im}(\eta)] = -2\textrm{Im}(\eta).
\end{align}
Hence, we have to maximize the negative imaginary part of the Bargmann invariant $\eta = \braket{abc}{cab}$. Since the operator is $U\otimes U\otimes U$-invariant, without loss of generality, we can set
\begin{align}
    \ket{a} &= \ket{0}\notag\\
    \ket{b} &= \cos(\theta_b) \ket{0}+e^{\mathbf{i}\varphi_b}\sin(\theta_b)\ket{\psi_b}\notag\\
    \ket{c} &= \cos(\theta_c) \ket{0}+e^{\mathbf{i}\varphi_c}\sin(\theta_c)\ket{\psi_c},
\end{align}
with $\braket{0}{\psi_b} = 0$ and $\braket{0}{\psi_c}=0$. This yields
\begin{align}
    -\textrm{Im}(\eta) &= -\textrm{Im}\left\{\cos(\theta_c)\cos(\theta_b)[\cos(\theta_b)\cos(\theta_c)+\sin(\theta_b)\sin(\theta_c)e^{\mathbf{i}(\varphi_b-\varphi_c)}\braket{\psi_b}{\psi_c}]\right\}\notag\\
    &= -\cos(\theta_c)\cos(\theta_b)\sin(\theta_b)\sin(\theta_c)r\sin(\varphi_c-\varphi_b+\phi),
\end{align}
where we wrote $\braket{\psi_b}{\psi_c} = re^{\mathbf{i}\phi}$. Clearly, this expression is maximized when $r=1$, $\sin(\varphi_c-\varphi_b+\phi)=-1$ and $\cos(\theta_b)\sin(\theta_b) = \cos(\theta_c)\sin(\theta_c) = \nicefrac{1}{2}$ and hence $\max [-\textrm{Im}(\eta)] = \nicefrac{1}{2}\times\nicefrac{1}{2}$, which coincides with the findings in \cite{ChaBarkmann, ChaBarkmann2}. Then we find:
\begin{align}
    \bra{abc}W_-\ket{abc} = \mathbf{i}\times d (\eta - \eta^*) \leq d\times 2\times \frac{1}{4} = \frac{d}{2}.
\end{align}
\end{proof}

Now, consider the separability bound of $W_+$ for the local dimension $d=3$. For dimension $d> 3$, the separability bound coincides with the biseparability bound, which will follow from the proof of Observation~4c below. 

\noindent
\textbf{Observation 4b.} \textit{
For the witness operator based on the anticommutator it holds $\tr(W_+\sigma) \leq \nicefrac{4}{3}$
for all fully separable states for local dimension $d=3$.
}
\begin{proof}
 Let us consider the expectation value of $W_+$ for a pure product state:
 \begin{align}
     \bra{abc}W_+\ket{abc} = \sum_{ijk=1}^8\kappa_{ijk}^+\bra{a}\lambda_i\ket{a} \bra{b}\lambda_j\ket{b} \bra{c}\lambda_k\ket{c} = \sum_{ij}g_{ij}\bra{a}\lambda_i\ket{a} \bra{b}\lambda_j\ket{b},
 \end{align}
 where $\{\lambda_i\}$ are the Gell-Mann matrices with $\tr(\lambda_i\lambda_j)=d\delta_{ij}$ and we defined $g_{ij} = \sum_k\kappa^+_{ijk}\bra{c}\lambda_k\ket{c}$. 
 Since the operator is $U^{\otimes 3}$-invariant, we can assume that $\ket{c} = \ket{2}$. Then $\bra{c}\lambda_k\ket{c}$ is nonzero if $\lambda_k$ has a nonzero entry $\ketbra{2}$, which is only the case for
 \begin{align}
     \lambda_8 = \left(\begin{matrix}
         \frac{1}{\sqrt{2}} & 0 & 0 \\
         0 & \frac{1}{\sqrt{2}} & 0\\
         0 & 0 & -\sqrt{2}
     \end{matrix}\right).
 \end{align}
 We can then calculate the matrix $g_{ij}$ and find that it has the form
 \begin{align}
     g = \frac{1}{3}\text{diag}(-2,1,1,-2,1,1,-2,2).
 \end{align}
 This yiels:
 \begin{align}
     \bra{abc}W_+\ket{abc} = \sum_{ij}g_{ij}\bra{a}\lambda_i\ket{a}\bra{b}\lambda_j\ket{b} \leq \frac{2}{3} \left|\sum_i \bra{a}\lambda_i\ket{a}\bra{b}\lambda_i\ket{b}\right| = \frac{2}{3}|\tr(f)|,
 \end{align}
where we defined $f_{ij}\coloneqq\bra{a}\lambda_i\ket{a}\bra{b}\lambda_j\ket{b}$. Since $f$ can be interpreted as the product of two Bloch vectors $f=\vec{a}\vec{b}^T$, with $a_i = \bra{a}\lambda_i\ket{a}$ and similar for $\vec{b}$, it is a rank-one projector and has only one nonzero eigenvalue $\iota$ and we have
 \begin{align}
     |\tr(f)| = \max \{0, |\iota|\} \leq |\vec{a}\cdot\vec{b}|\leq |\vec{a}||\vec{b}| = 2,
 \end{align}
where in the last step we used that the norm of a bloch vector is given by $\sqrt{d-1}$.

Then we can complete the proof:
\begin{align}
    \bra{abc}W_+\ket{abc} \leq \frac{2}{3}\times2 = \frac{4}{3}.
\end{align}
Note that following this calculation the bound is saturated when also $\ket{a} = \ket{b} = \ket{2}$. Due to the $U^{\otimes 3}$-invariance, it follows more general that the bound is saturated when $\ket{a} = \ket{b} = \ket{c}$.
\end{proof}
Next, let us consider the biseparability bound of the operator $W_+$. From this proof also follows the separability bound for the local dimension $d>3$.

\noindent
\textbf{Observation 4c.}\textit{
For the witness operator based on the anticommutator it holds
$\tr(W_+\sigma) \leq [2(d-1)(d-2)]/d$
for all fully separable and for all biseparable states of local dimension $d>3$.}

\textit{For all biseparable states of local dimension $d=3$ it further holds $\tr(W_+\varrho_{\mathrm{bisep}}) \leq \nicefrac{10}{3}$.
}
\begin{proof}
Let us start with the biseparability bound; we will then see that for dimension $d>3$ it coincides with the separability bound. 
Since the operator $W_+$ is invariant under $U\otimes U\otimes U$-transformations, we can assume that the biseparable state maximizing the expectation value $\bra{\varphi_\mathrm{bisep}}W_+\ket{\varphi_\mathrm{bisep}}$ is given by $\ket{0}_A\ket{\mu}_{BC}$. More precisely, we would have to consider also the other bipartitions $\ket{0}_B\ket{\mu}_{AC}$ and $\ket{0}_C\ket{\mu}_{AB}$. However, in this case all bipartitions give the same reduced operator $\bra{0}_AW_+\ket{0}_A$.
It holds:
\begin{align}
    \max_{\ket{\varphi_\mathrm{bisep}}}\bra{\varphi_\mathrm{bisep}}W_+\ket{\varphi_\mathrm{bisep}} = \max_{\ket{\mu}}\bra{0}_A\bra{\mu}_{BC}W_+\ket{0}_A\ket{\mu}_{BC} = \max \text{Eig}(\bra{0}_AW_+\ket{0}_A)
\end{align}
So, to compute the the biseparability bound, we only need to find the largest eigenvalue of the reduced operator $Y^+_{BC}=\bra{0}_AW_+\ket{0}_A$.

Consider the witness operator's decomposition into permutation operators (Eq.~(\ref{eq:decWS})). Then, we can compute
\begin{align}
    Y^+_{BC} &= \bra{0}_A\left(d(T+T^2)-2(F_{12}+F_{23}+F_{13})+\frac{4}{d}\mathds{1}\right)\ket{0}_A\notag\\
    &=d\left(\sum_i\ket{0i}\bra{i0}+\sum_i\ket{i0}\bra{0i}\right)-2\left(\sum_i\ketbra{0i}+\sum_i\ketbra{i0}+\sum_{ij}\ket{ij}\bra{ji}\right)+\frac{4}{d}\sum_{ij}\ketbra{ij}.
\end{align}
The eigenvectors and their corresponding eigenvalues read
\begin{align}
    \lambda_1 &= \frac{4}{d}-2,\,\,e^{ij}_1= \ket{ij}+\ket{ji},\,\,\text{for}\,\, 0\neq i\neq j \neq 0\\
    \lambda_2 &= \frac{4}{d}+2,\,\,e_2^{ij}=\ket{ij}-\ket{ji},\,\,\text{for}\,\, 0\neq i\neq j \neq 0\\
    \lambda_3 &= \frac{4}{d}+d-4,\,\,e_3^{i0}=\ket{i0}+\ket{0i},\,\,\text{for}\,\,i\neq 0\\
    \lambda_4 &= \frac{4}{d}-d,\,\,e_4^{i0} = \ket{i0}-\ket{0i},\,\,\text{for}\,\,i\neq 0\\
    \lambda_5 &= \frac{4}{d}-2,\,\,e_5^i=\ket{ii},\,\,\text{for}\,\,i\neq 0\\
    \lambda_6 &= \frac{4}{d}+2d-6,\,\,e_6 = \ket{00}
\end{align}
The only two options for the largest eigenvalue are $\lambda_2$ and $\lambda_6$. We start considering $d=3$. Then we have $\lambda_2 = \nicefrac{10}{3}$ and $\lambda_6 = \nicefrac{4}{3}$, and so the largest eigenvalue is $\lambda_2 =\nicefrac{10}{3}$. It follows:
\begin{align}
    \tr(W_+\varrho_\mathrm{bisep}) \leq \frac{10}{3}
\end{align}
for all biseparable three-qutrit states.

For local dimension $d>3$ we find that always $\lambda_6\geq \lambda_3$ holds. But since the eigenvector corresponding to $\lambda_6$ is fully separable, this bound also holds for full separability and we have:
\begin{align}
    \tr(W_+\sigma)\leq\frac{4}{d}+2d-6=\frac{2(d-1)(d-2)}{d}.
\end{align}
for all fully separable and biseparable states $\sigma$ in local dimension $d>3$.
\end{proof}
Since also $W_-$ is $U\otimes U\otimes U$-invariant, we can proceed analogously to show its biseparability bound.

\noindent
\textbf{Observation 4d.}\textit{
For the witness operator based on the commutator it holds $\tr(W_-\varrho_{\mathrm{bisep}}) \leq d$
for all biseparable states.
}
\begin{proof}
    With the same argument as above, it suffices to consider the largest eigenvalue of the reduced operator $Y_{BC}^-$. We have
    \begin{align}
        Y_{BC}^-= \mathbf{i}\times d\left(\sum_i\ket{0i}\bra{i0}-\ket{i0}\bra{0i}\right)
    \end{align}
    and the eigenvalues and eigenvectors are:
    \begin{align}
        \lambda_1 &= -d,\,\,e_1=\ket{0i}+\mathbf{i}\ket{i0},\,\,\text{for}\,\,i\neq 0\\
        \lambda_2 &= d,\,\,e_2=\ket{0i}-\mathbf{i}\ket{i0},\,\,\text{for}\,\,i\neq 0\\
        \lambda_3 &= 0,\,\, e_3 =\ket{ij},\,\,\text{for}\,\,i,j\neq 0\,\,\text{or}\,\,i=j=0 
    \end{align}
    Hence, the largest eigenvalue is given by $d$ and we have:
    \begin{align}
        \tr(W_-\varrho_{\mathrm{bisep}})\leq d
    \end{align}
    for all biseparable states.
\end{proof}
Lastly, the quantum bounds are given by the largest eigenvalues of the operators. So, from the main text we know:
\begin{align}
    \langle W_-\rangle &\leq d\sqrt{3}\\
    \langle W_+\rangle &\leq \frac{2}{d}(d+1)(d+2),
\end{align}
where the chiral and antichiral symmetric states saturate the quantum bound of $W_-$ and the antisymmetric states saturate the quantum bound of $W_+$.

\section{Details on Figure~1 in the main text}
\begin{figure}[htbp]
    \centering
    \includegraphics[width=0.48\linewidth]{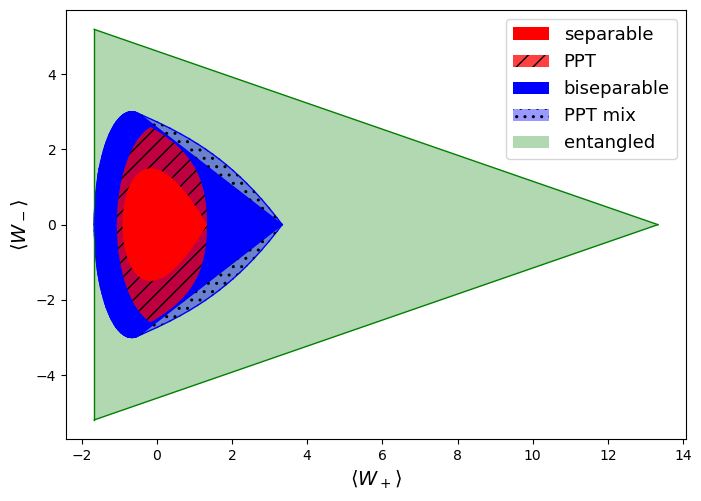}
    \includegraphics[width=0.43\linewidth]{UUUexpvals4new.png}
    \caption{Three-qutrit (left) and three-ququad (right) state space, parametrized by the expectation values of $W_\pm$. We differ between states that are fully separable (red), PPT with respect to all bipartitions (red hatched), biseparable (dark blue), mixtures of PPT states (red, dark blue and blue dotted) and GME (blue dotted and green). Note that the state spaces form a polytope since the witness observables commute. On the vertices of the triangle we find the chiral, antichiral and antisymmetric states, respectively. The symmetric states are lying on the vertex of the fully separable area. Note that in the qutrit plot we have an AME state on each vertex of triangle. We further find that combining the witnesses $W_\pm$, we can detect PPT entangled states. In fact, in the ququad plot, the red hatched area intersecting with the blue dotted area uncovers $U^{\otimes 3}$-invariant PPT GME states. Note, that although for the qutrit plot such an area is not visible, we will see later that PPT GME is also present for local dimension $d=3$.}
    \label{fig:statespaceUUU1}
\end{figure}

In this appendix we give more detail on Fig.~1 in the main text.
In Fig.~\ref{fig:statespaceUUU1} we plot the state space for three qutrits and three ququads (see also main text), parametrized by the expectation values of the operators $W_\pm$.
To obtain the bounds for full separablity and biseparability, we compute
\begin{align}
   \max_{\ket{\varphi}} \bra{\varphi}\cos(\theta) W_-+\sin(\theta) W_+\ket{\varphi},
\end{align}
for $\ket{\varphi}\in \mathrm{SEP}$ and $\ket{\varphi}\in \mathrm{BISEP}$, respectively, and $\theta \in[0,2\pi]$. To do so, we use an algorithm, similar to the one in Ref~\cite{SteinbergGuehne2024}. By taking the largest eigenvalues of the combined witnesses we obtain the quantum bounds.
The red PPT area is the area of states for which the partial transpose with respect to all bipartitions is positive (PPT). These are obtained by a semidefinite programme (SDP):
\begin{align}
    \min &\{- \tr[\varrho (\cos(\theta) W_-+\sin(\theta) W_+)]\}\notag\\
    \text{s.\,\,th.}\,\,  &\varrho \geq 0,\,\,\tr(\varrho)=1,\notag\\
    &\varrho^{T_A} \geq 0,\notag\\
    &\varrho^{T_B} \geq 0,\notag\\
    &\varrho^{T_C} \geq 0.
\end{align}
Similarly, the blue PPT mix area denotes the states that can be written as a convex combination of states that are PPT with respect to different bipartitions \cite{JungnitschMoroderGuehne2011}, i.e.
\begin{align}
    \varrho^{\text{PPT\,mix}} = p_1 \varrho^{\mathrm{PPT}\,A|BC} + p_2 \varrho^{\mathrm{PPT}\,B|AC}+p_3\varrho^{\mathrm{PPT}\,C|AB},
\end{align}
with $p_1+p_2+p_2=1$ and $p_i\geq 0$ for $i=1,2,3$.
We obtain this area by running the SDPs
\begin{align}
    \min &\{- \tr[\varrho (\cos(\theta) W_-+\sin(\theta) W_+)]\}\notag\\
    \text{s.\,\,th.}\,\,  &\varrho \geq 0,\,\,\tr(\varrho)=1,\notag\\
    &\varrho^{T_i} \geq 0,
\end{align}
for $i=A,B,C$ and taking the convex hull of the three areas obtained by that.

Now, let us discuss the results. There are three points to mention.

\noindent
\textbf{Remark 1.} The set of entangled states is a polytope (a triangle). This is clear since the operators $W_\pm$ commute and therefore have common eigenvectors. In fact, we can also directly find the boundary of the quantum area by considering the largest eigenvalues of the operators $W_\pm$ and taking linear combinations of the corresponding eigenvectors.

\noindent
\textbf{Remark 2.} The states on the vertices of the triangle are the fully antisymmetric state(s) at the maximum of $\langle W_+\rangle$ and the (anti)chiral symmetric states at the maximum and minimum of $\langle W_-\rangle$. In particular, note that for local dimension $d=3$ all states on the vertices are absolutely maximally entangled (AME).

\noindent
\textbf{Remark 3.} The witness operator $W_+$ and especially the combination of $W_+$ and $W_-$ can detect PPT-entangled states. For three ququads we even detect states, which are PPT for all bipatitions and still genuine multipartite entangled (GME). We will have a closer look at this in the next appendix.

\section{Proof of Observation 5}
We will first show that the operators $P$ and $\bar{P}$ defined in Observation~5 are entanglement witnesses for genuine multipartite entanglement. Then we will discuss the PPT GME states detected by these witnesses.
\subsection{The witnesses}

\noindent
\textbf{Observation 5.}\textit{
The operators $P = \Pi_\mathcal{J}-\Pi_\mathcal{A}$ and $\bar{P}=\Pi_{\bar{\mathcal{J}}}-\Pi_\mathcal{A}$,
are entanglement witnesses for genuine multipartite entanglement.
}
\begin{proof}
    Note that $\bar{P} = P^*$, so the proofs will be analogous and we now focus on the witness $P$.
    Since $P$ is $U\otimes U\otimes U$-invariant, we can assume that the expectation value is maximized by a state of the form $\ket{\varphi^{A|BC}} = \ket{0}_A\ket{\mu}_{BC}$.\footnote{Note that as above fixing Bob's qubit yields the hermitian conjugated reduced operator, while fixing Charlie's qubit yields the same.} Consider the reduced operator:
    \begin{align}
        \bra{0}_A6P\ket{0}_A &= \bra{0}_A \left(\mathds{1}+(2\omega-1)T^2+(2\omega^2-1)T+F_{12}+F_{23}+F_{13}\right)\ket{0}_A\notag\\
        &= \sum_{ij}\ket{ij}\bra{ij} + (2\omega -1)\sum_i \ket{i0}\bra{0i}+(2\omega^2-1)\sum_i \ket{0i}\bra{i0}+\sum_i \ket{0i}\bra{0i}+\sum_{ij} \ket{ij}\bra{ji}+\sum_i \ketbra{i0} \coloneqq Y_{BC}
    \end{align}
    We will now show that this is a positive operator, which implies that $P$ has a positive expectation value for all biseparable states and therefore is a witness for GME.
    
    To get an idea how the proof works we consider the structure of the operator $Y_{BC}$ for $d=3$, however we will give a proof valid for all dimensions:
    \begin{align}
        Y_{BC} = \begin{pNiceArray}{ccc|ccc|ccc}
             0 &   &   &   & & &   & &  \\
               & 2 &   & 2\omega & & &   & & \\
               &   & 2 &   & & & 2\omega & & \\
            \hline
               & 2\omega^* &   & 2 &   &  &   &   &  \\
               &   &   &   & 2 &  &   &   & \\
               &   &   &   &   & 1 &  & 1 & \\
            \hline
               &   & 2\omega^* &   &   &  &  2 &   &  \\
               &   &   &   &   & 1 &   & 1 & \\
               &   &   &   &   &  &  &  & 2\\
        \end{pNiceArray}.
    \end{align}
    Let us clarify the structure in more general terms:
    The entry $\ketbra{00}$ is given by $2+2\omega+2\omega^* = 0$. All further entries on the diagonal (excluding the case $i=j=0$) are given by $\ket{ij}\bra{ij} = 1+\delta_{i0} + \delta_{0j} + \delta_{ij}$. The only off-diagonal terms are $\ket{ij}\bra{ji} = (2\omega-1)\delta_{i0} + (2\omega^*-1)\delta_{0j} +1$.

    To verify that $Y_{BC}$ is a positive semidefinite matrix, we have to show that all its main minors are nonnegative. First, let us clarify what the main minors are:

    \noindent
    \textbf{Definition (see also Ref.~\cite{BurnsidePanton2017}).} \textit{Main minors are the determinants of all submatrices of a matrix $M$ obtained by crossing out rows and columns of $M$ such that the crossed rows and columns meet at the diagonal.}
    \\ 
    The determinant of a $k\times k$-submatrix $M_k$ is called main minor of the order $k$. For example, all diagonal entries of $M$ are main minors of order one.
    
    Let us take a precise look at the main minors of $Y_{BC}$ now. First, we easily see that all diagonal entries are nonnegative, so all main minors of order one fulfill the criterion and we can continue considering the main minors of order $k\geq 2$. 
    
    Most of the submatrices are diagonal matrices with positive entries on the diagonal, so we immediately see that their determinant is nonnegative as well. Hence, we only need to take a look at the nondiagonal submatrices. 
    
    We now give an algorithm to construct these. Then, we will see that all submatrices corresponding to the main minors are at most either block diagonal with one nondiagonal $2\times2$ block or have a shape such that the determinant will reduce to a product of nonnegative diagonal entries and the determinant of a $2\times2$ block. The algorithm is as follows:
    \begin{itemize}
        \item[1.] Pick a row (and the respective column, such that they meet at the diagonal) that has nonzero entries on the off-diagonal.
        \item[2.] In that row, pick the column (and its respective row) corresponding to the nonzero entry on the off-diagonal.
        \item[3.] In the previously chosen row, pick another column (and its respective row) with a nonzero entry on the off-diagonal.
        \item[4.] Repeat until $k$ rows and their respective columns are chosen, all other rows and columns are crossed out.
    \end{itemize}

    To find the main minors of second order, we start choosing a row $\bra{ij}$ with $i\neq j$. Since the operator $Y_{BC}$ is symmetric, the second row we pick is $\bra{ji}$. Then we obtain\footnote{Note that in this notation the case $i=j=0$ is excluded}
    \begin{align}
        \det(M_2) = \left|\begin{matrix}
            1+\delta_{i0} + \delta_{0j}  & (2\omega-1)\delta_{i0} + (2\omega^*-1)\delta_{0j} +1\\
            (2\omega-1)\delta_{j0} + (2\omega^*-1)\delta_{0i} +1 & 1+\delta_{j0} + \delta_{0i} 
        \end{matrix}\right| = (1 +3\delta_{j0}+3\delta_{i0}) - (3\delta_{i0}+3\delta_{j0}+1) = 0
    \end{align}
    and it follows that all main minors of second order are nonnegative as well. 
    
    To construct the third order main minors, we start in the same way, however there are no rows in $Y_{BC}$ containing more than two nonzero entries. Therefore, the submatrix will be either blockdiagonal where one block is the submatrix $M_2$, for example 
    \begin{align}
        M_3 = \det\begin{pNiceArray}{cc|c}
         \Block{2-2}<\Large>{M_2} && 0\\
                                         &&  0\\\hline
                             0       &  0    & a 
        \end{pNiceArray} = a\det(M_2) = 0.
    \end{align}
    or have a shape such that crossing out the row and column with nonzero off-diagonal entries will always yield $M_2$
        \begin{align}
        M_3 = \det\begin{pNiceArray}{ccc}
         M_2^{00} & 0& M_2^{01}\\
                    0  & a&  0\\
        M_2^{10}    &  0    & M_2^{11} 
        \end{pNiceArray} = a\det(M_2) = 0.
    \end{align}
    This completes the proof since for all bigger submatrices, more precise the main minors of order $k\geq 4$ the same argument holds and we can only find main minors which reduce to a product of the nonnegative diagonal terms and the determinant of the matrix $M_2$.    
\end{proof}
Note, that following a short calculation we can see that the diagonal border of the PPT GME area for ququads in Fig.~\ref{fig:statespaceUUU1} is described by $P = \Pi_\mathcal{J}-\Pi_\mathcal{A}$ for positive $\langle W_-\rangle$. For negative $\langle W_-\rangle$ is given by $\bar{P}$.\\

\subsection{Discussion of the PPT GME states}
Now, let us characterize the PPT GME states that are detected by the witness $P$ and note again that for $\bar{P}$ we can proceed analogously since $\bar{P}=P^*$.

From the semidefinite programme (SDP) we find that the states are of the following form:
\begin{align}
    \varrho = a\Pi_\mathcal{A}+b\Pi_\mathcal{S}+c\Pi_\mathcal{\Bar{J}}.\label{eq:PPTGME}
\end{align}
We now derive the conditions for the coefficients $a,b,c$ such that $\varrho$ is a PPT GME state.

Let us start with the conditions for which $\varrho$ is a quantum state. First, note that the above expression is already a spectral decomposition of $\varrho$. Hence, its eigenvalues are given by $a,b$ and $c$ and it follows the first condition:
\begin{align}
    a,b,c\in \mathbb{R},\,a\geq 0,\,\,b\geq 0,\,\, c\geq 0.
\end{align}
For $\varrho$ to be normalized, we require
\begin{align}
    1 = \tr(\varrho) = a\tr(\Pi_\mathcal{A})+b\tr(\Pi_\mathcal{S})+c\tr(\Pi_\mathcal{\Bar{J}}) 
    \Rightarrow b = \frac{1-a\tr(\Pi_{\mathcal{A}})-c\tr(\Pi_{\mathcal{\Bar{J}}})}{\tr(\Pi_{\mathcal{S}})} \coloneqq f(a,c),
\end{align}
with
\begin{align}
    \tr(\Pi_\mathcal{A}) = \frac{1}{6}(d^3+2d-3d^2),\,\,
    \tr(\Pi_\mathcal{S}) = \frac{1}{6}(d^3+2d+3d^2),\,\,
    \tr(\Pi_\mathcal{\Bar{J}}) = \frac{1}{3}(d^3-d).
\end{align}

Now, we derive the condition for which $\varrho$ is GME, or more precise, for which it is detected by $P$. We have:
\begin{align}
    &0>\tr(P\varrho) = -a\tr(\Pi_{\mathcal{A}}) \Rightarrow a > 0
\end{align}
Note that this means that the state of the form in Eq.~(\ref{eq:PPTGME}) is GME as soon as it has antisymmetric states in its spectrum.

\begin{figure}[t]
    \centering
    \includegraphics[width=0.45\linewidth]{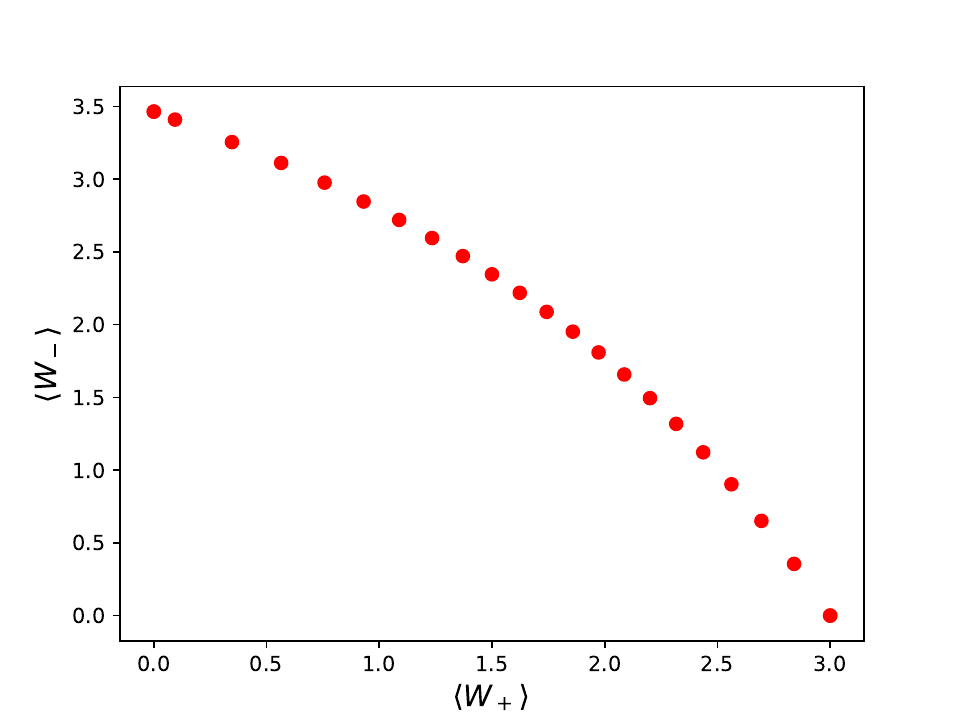}
    \includegraphics[width=0.45\linewidth]{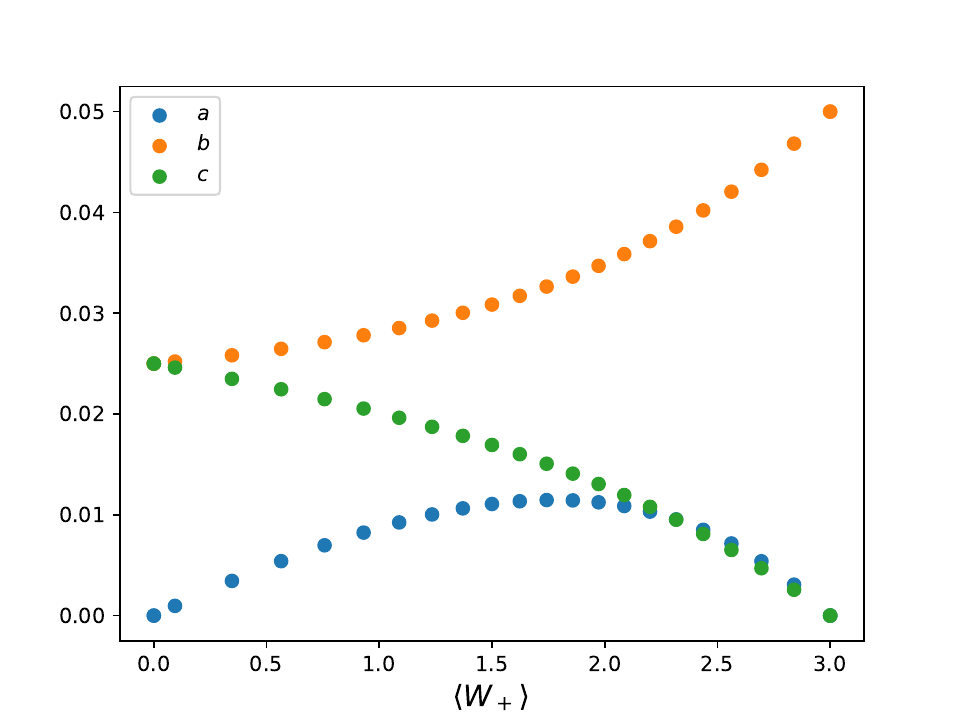}
    \caption{PPT GME for three ququads: Left: The states $\varrho (a,b,c)$ that saturate bound for PPT GME, i. e. all states above are GME but not PPT for all bipartitions, in terms of the expectation values $\langle W_\pm\rangle$ (compare also Fig.~\ref{fig:statespaceUUU1}). Right: The coefficients $a,b,c$ corresponding to the states in the left figure in terms of $\langle W_+\rangle$.}
    \label{fig:coeffsPPTGME}
\end{figure}

Lastly, we want to find the conditions for which $\varrho$ has a positive partial transpose for all bipartitions. Although we can in principle write down $\varrho^{T_i}(a,f(a,c),c)$ and find the eigenvalues, there is no compact condition of the form $a = g(c)$ such that the smallest eigenvalues for each bipartition are nonnegative. Therefore, we make use of the SDP again. In particular, we consider the upper border of the PPT GME area for local dimension $d=4$ (see also Fig.~\ref{fig:coeffsPPTGME} (left)). In Fig.~\ref{fig:coeffsPPTGME} (right) we plot the coefficients $a,b$ and $c$ with which the state $\varrho$ lies on the upper border of the PPT area (see also Fig.~\ref{fig:statespaceUUU1}).

Interestingly, the coefficients $b$ and $c$ seem to be mirrored on the axis $y=0.01$, while the coefficient $a$, corresponding to the antisymmetric subspace grows until a certain maximum and decreases again to the value zero. Indeed, the first point corresponds to a $\varrho_0 \propto \Pi_\mathcal{S}+\Pi_\mathcal{\bar{J}}$ which is not detected by $P$ since its antisymmetric part is zero and the last point to the fully separable state $\varrho_n \propto \Pi_\mathcal{S}$.

By this, we characterized the three-ququad states that lie in the upper PPT GME area in Fig.~\ref{fig:statespaceUUU1}. They can be detected by the witness $P$ as well as by combining $W_\pm$. There are three more remarks to mention.

    \noindent
\textbf{Remark 1.} Considering the PPT GME area in Fig.~\ref{fig:statespaceUUU1} where $\langle W_-\rangle<0$, the witness to decide GME is given by $\bar{P}=\Pi_\mathcal{\bar{J}}-\Pi_\mathcal{A}$ and the states on the outer border will be of the form $\bar{\varrho}=a\Pi_\mathcal{A}+b\Pi_\mathcal{S}+c\Pi_\mathcal{J}$, which is simply the complex conjugation.

\begin{figure}[b]
    \centering
    \includegraphics[width=0.45\linewidth]{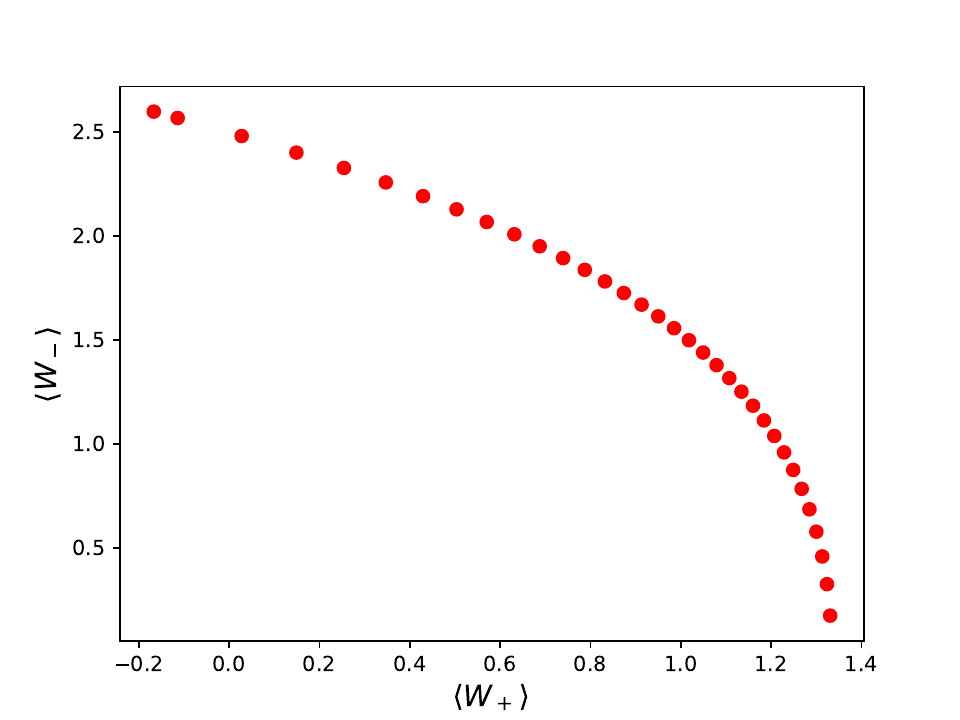}
    \includegraphics[width=0.45\linewidth]{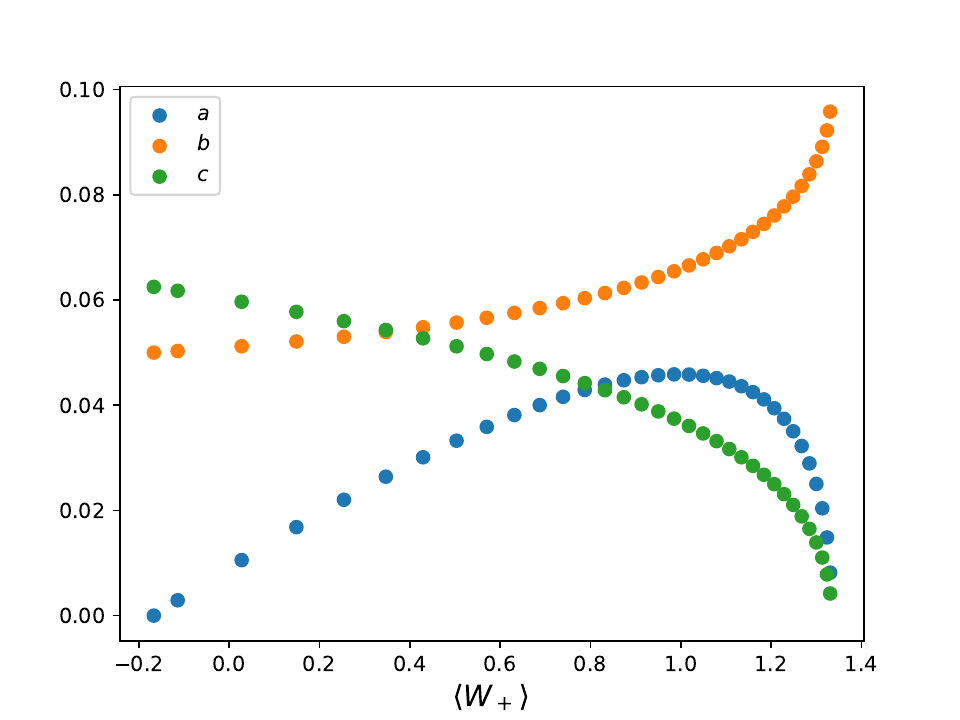}
    \caption{PPT GME for three qutrits: Left: The states $\varrho (a,b,c)$ that saturate bound for PPT GME, i. e. all states above are GME but not PPT for all bipartitions, in terms of the expectation values $\langle W_\pm\rangle$ (compare also Fig.~\ref{fig:statespaceUUU1}). Right: The coefficients $a,b,c$ corresponding to the states in the left figure in terms of $\langle W_+\rangle$.}
    \label{fig:coeffsPPTGME3}
\end{figure}

    \noindent
\textbf{Remark 2.} Although in Fig.~\ref{fig:statespaceUUU1} there is no such PPT GME area for local dimension $d=3$, we still find that they exist. As shown before, the witness $P$ is valid for all dimensions $d\geq 3$, so the only condition needed for $\varrho(a,b,c)$ to be detected as GME is that $a>0$. Proceeding analogously to the ququad case we find the results shown in Fig.~\ref{fig:coeffsPPTGME3}, so indeed all the states plotted there (besides the very first one) are PPT GME as well.

    \noindent
\textbf{Remark 3.} In Ref~\cite{EggelingWerner2001} is was shown that for $U\otimes U\otimes U$-invariant three-qubit states, biseparability is equivalent to the positivity of the partial transpose, so there cannot exits PPT GME states for $d=2$.

\section{Discussion of the $U^*\otimes U\otimes U$-invariant witness observable $\mathbb{W}_\pm$ and further generalizations}
In the main text, we defined the witness observable $\mathbb{W}_- = W_-^{T_1}$ in order to construct the flip-conjugate subspace from its eigenvectors. We will now discuss further properties and generalizations thereof. First, note that we can analogously define $\mathbb{W}_+ = W_+^{T_1}$, so we have
\begin{align}
    \mathbb{W}_\pm = W_\pm^{T_1}.
\end{align}
These are linear combinations of partially transposed permutations and therefore $U^*\otimes U\otimes U$-invariant \cite{Weyl1946,EggelingWerner2001}. Note that analogously, one can define operators $W_\pm^{T_2}$ and $W_\pm^{T_3}$, which are $U\otimes U^*\otimes U$- and $U\otimes U\otimes U^*$-invariant, respectively, but in the following we focus on $\mathbb{W}_\pm = W_\pm^{T_1}$.

    \noindent
\textbf{Proposition 3.}\textit{
The operators $\mathbb{W}_\pm$ can be implemented via
\begin{align}
    \mathbb{W}_- &= \sum_{ijk}-\frac{\mathbf{i}}{d^2}\tr((\lambda_i\otimes [\lambda_j,\lambda_k]) F^{T_1})\lambda_i\otimes\lambda_j\otimes\lambda_k\label{eq:newopm}\\
    \mathbb{W}_+ &= \sum_{ijk}\frac{1}{d^2}\tr((\lambda_i\otimes \{\lambda_j,\lambda_k\}) F^{T_1})\lambda_i\otimes\lambda_j\otimes\lambda_k.\label{eq:newopp}
\end{align}
}
Note, that we applied the swap-trick $\tr((A\otimes B) F)=\tr(AB)$ \cite{EltschkaSiewert2018} to the structure constants $\kappa^\pm_{ijk}$ to construct $\mathbb{W}_\pm$ from $W_\pm$.
\begin{proof}
    It is sufficient to consider the reduced operator $\mathbb{W}$. Then we have
\begin{align}
    \mathbb{W} &= \sum_{ijk}\tr[(\lambda_i \otimes \lambda_j\lambda_k)F^{T_1})\lambda_i\otimes\lambda_j\otimes\lambda_k = \sum_{ijk}\tr[(\lambda_i^T \otimes \lambda_j\lambda_k)F)\lambda_i\otimes\lambda_j\otimes\lambda_k\notag\\
    &= \sum_{\lambda_i\,\,\mathrm{real}}\sum_{jk} \tr[(\lambda_i \otimes \lambda_j\lambda_k)F)\lambda_i\otimes\lambda_j\otimes\lambda_k - \sum_{\lambda_i\,\,\mathrm{imag}}\sum_{jk} \tr[(\lambda_i \otimes \lambda_j\lambda_k)F)\lambda_i\otimes\lambda_j\otimes\lambda_k\notag\\
    &= \sum_{ijk}\tr[(\lambda_i \otimes \lambda_j\lambda_k)F)\lambda_i^T\otimes\lambda_j\otimes\lambda_k = W^{T_1},
\end{align}
where in the second equation we used that all Gell-Mann matrices are either completely real or completely imaginary, so transposing either gives a sign or leaves the matrix unchanged. 
\end{proof}

We want to give a few remarks.
    
\begin{figure}[htbp]
    \centering
    \includegraphics[width=0.45\linewidth]{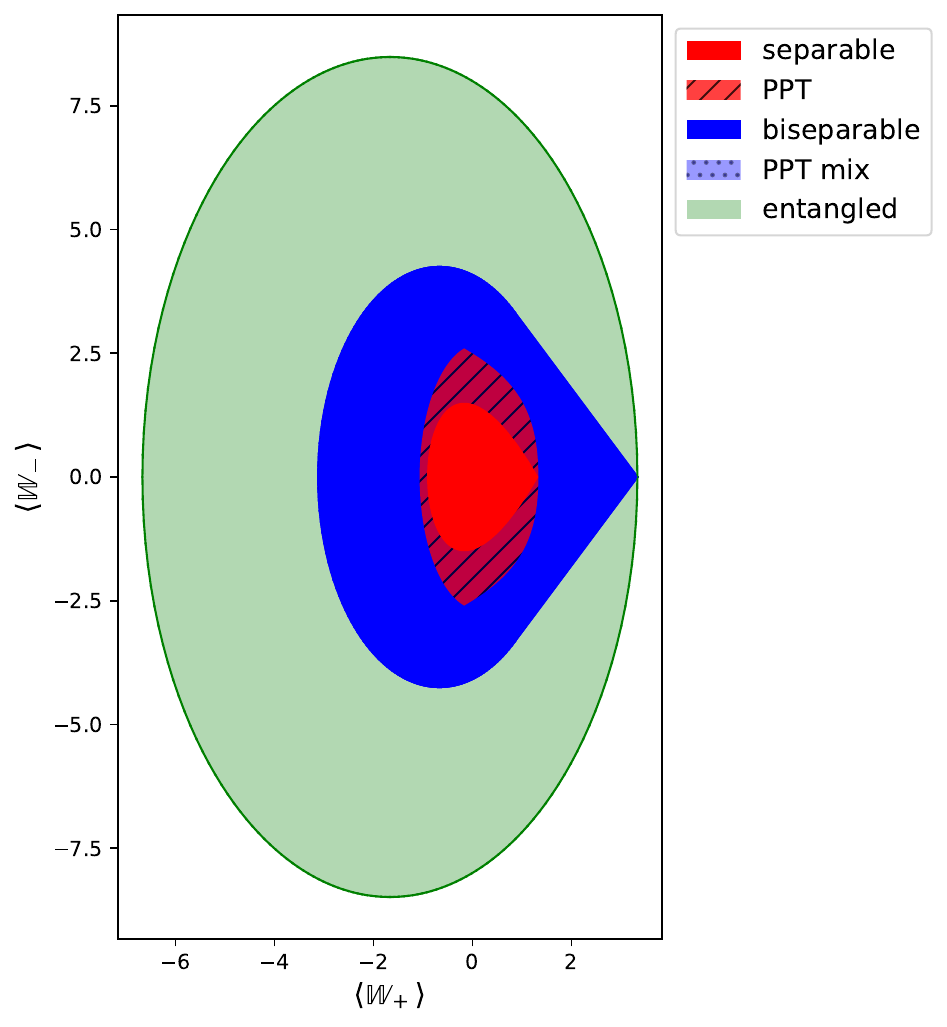}
    \includegraphics[width=0.45\linewidth]{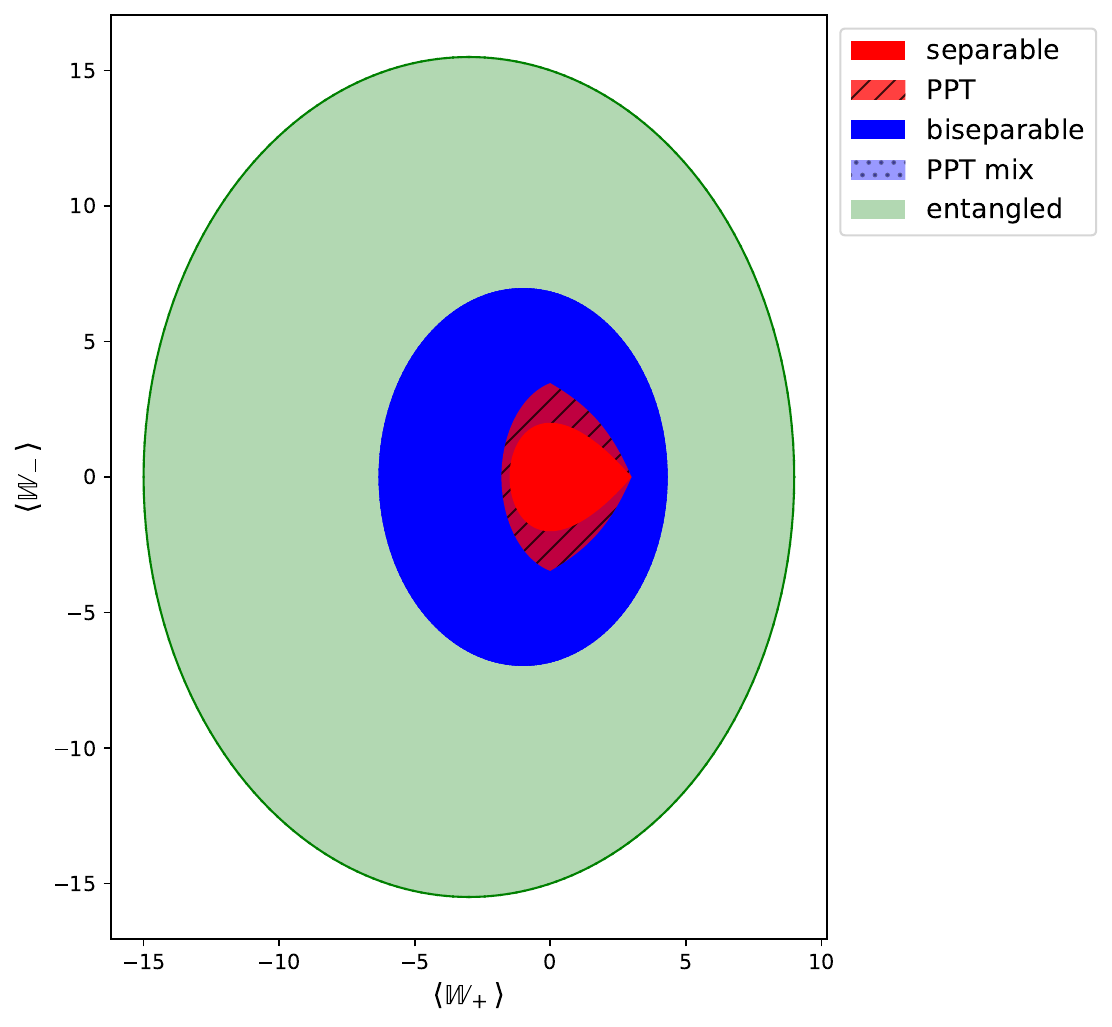}
    \caption{Three-qutrit (left) and three-ququad (right) state space, parametrized by the expectation values of $\mathbb{W}_\pm$. We differ between states that are fully separable (red), PPT for all bipartitions (red hatched), biseparable (dark blue) and GME (green). The area that contains states which are a mixture of PPT states coincides with the biseparable area. Note that, due to the symmetry breaking of the partial transpose, for the operators $\mathbb{W}_\pm$ the expectation values for biseparable states depend on the bipartition (different from the operators $W_\pm$). Hence, the biseparable and PPT areas are shaped differently than in Fig.~\ref{fig:statespaceUUU1}. Moreover, since the operators $\mathbb{W}_\pm$ do not commute, here, the state space is not a polytope.}
    \label{fig:statespaceOOO}
\end{figure}

    \noindent
\textbf{Remark 1.} Similar to the operators $W_\pm$ we can also plot the state space in terms of expectation values of $\mathbb{W}_\pm$, which is shown in Fig.~\ref{fig:statespaceOOO}. Different from the $U\otimes U\otimes U$-invariant case, the states space is no polytope, since $\mathbb{W}_\pm$ do not commute and therefore do not have common eigenvectors. We further see that with combinations of $\mathbb{W}_+$ and $\mathbb{W}_-$ one can detect states that are PPT for all bipartitions and biseparable but not GME. Moreover, states that are a mixture of PPT states are not detected.

    \noindent
\textbf{Remark 2.} The operators can be used to define another highly entangled subspace. While the eigenvectors of $W_-$ are given by the antichiral and chiral symmetric basis vectors, the eigenvectors of $\mathbb{W}_-$ also span the flip-conjugate symmetric subspace, corresponding to $F_{23}$. Note that respectively, $W_-^{T_2}$ corresponds to $F_{13}$ and $W_-^{T_3}$ to $F_{12}$.
    
    \noindent
\textbf{Remark 3.} We can generalize this construction replacing $F^{T_1}$ in Eq.~(\ref{eq:newopm}) and Eq.~(\ref{eq:newopp}) by an operator $M$ that is invariant under a subgroup of the unitary group (e.g. $O\otimes O M O^T\otimes O^T=M$). The observables obtained by this inherit the invariance and will be $O\otimes O \otimes O$-invariant. It is open for future work to investigate whether these operators uncover interesting subspaces as well.

In the following we briefly show, how the invariance is inherited. 

Let us consider the reduced operator $\mathbb{W}$ again. We now perform the computation for orthogonal operators $O$, however, note that this is analogous for any subgroup of the unitary group.
\begin{align}
    O \otimes O \otimes O \mathbb{W} O^T \otimes O^T \otimes O^T &=  \sum_{ijk}\tr[(\lambda_i\otimes \lambda_j\lambda_k)M] O\lambda_iO^T\otimes O\lambda_j O^T \otimes O\lambda_k O^T\notag\\
    &= \sum_{ijk}\tr[(O\otimes O)(\lambda_i\otimes \lambda_j \lambda_k) (O^T \otimes O^T)(O\otimes O)M(O^T \otimes O^T)] O\lambda_iO^T \otimes O\lambda_j O^T \otimes O\lambda_k O^T\notag\\
    &= \sum_{ijk}\tr[(O\lambda_iO^T\otimes O\lambda_j \lambda_kO^T)M] O\lambda_iO^T\otimes O\lambda_jO^T \otimes O\lambda_kO^T\notag\\
    &= \sum_{ijk}\tr[(O\lambda_iO^T\otimes O\lambda_jO^T O \lambda_k O^T)M] O\lambda_iO^T\otimes O\lambda_jO^T \otimes O\lambda_kO^T,
\end{align}
where we used $OO^T = O^TO =\mathds{1}$ and the $O\otimes O$-invariance of $M$. Then, similar to above, we rewrite $O\lambda_iO^T = \sum_\alpha \tilde{O}_{i\alpha} \lambda_\alpha$, which yields
\begin{align}
 &O \otimes O \otimes O \mathbb{W} O^T \otimes O^T \otimes O^T = \sum_{ijk} \tr[(\sum_\alpha \tilde{O}_{i \alpha} \lambda_\alpha\otimes \sum_\alpha \tilde{O}_{j \beta} \lambda_\beta  \sum_\gamma \tilde{O}_{k \gamma} \lambda_\gamma)M]
  \sum_{\alpha '}\tilde{O}_{i\alpha '}\lambda_{\alpha '}\otimes \sum_{\beta '}\tilde{O}_{j\beta '}\lambda_{\beta '} \otimes \sum_{\gamma '}\tilde{O}_{k\gamma '}\lambda_{\gamma '}\notag\\
  = &\sum_{\alpha \beta \gamma} \sum_{\alpha ' \beta ' \gamma '} \delta_{\alpha \alpha '} \delta_{\beta \beta '} \delta_{\gamma \gamma '}\tr[(\lambda_\alpha \otimes \lambda_\beta \lambda_\gamma) M] \lambda_{\alpha '}\otimes \lambda_{\beta '} \otimes \lambda_{\gamma '}
  = \sum_{ijk} \tr[(\lambda_i \otimes \lambda_j\lambda_k)M] \lambda_i \otimes \lambda_j \otimes \lambda_k = \mathbb{W}.
\end{align}
We find that replacing $F^{T_1}\mapsto M$, with $O\otimes O M O^T\otimes O^T = M$, yields operators that are invariant under $O\otimes O \otimes O$-transformations as well as independent of the choice of the constructing operators $\{\lambda_i\}$. 

\end{document}